\documentclass[longauth]{aa} 

\newcommand{\Nstars}{HIP~65A, TOI-157 and TOI-169}
\newcommand{\Nplanets}{HIP~65Ab, TOI-157b, and TOI-169b}

\providecommand{\feh}{$\left[{\rm Fe / H}\right]$}
\newcommand{\kms}{km\,s$^{-1}$}

\newcommand{\ms}{m\,s$^{-1}$}

\newcommand{\masy}{mas\,yr$^{-1}$}
\newcommand{\mpl}{M$_{p}$}

\newcommand{\teff}{$T_{\rm eff}$}
\newcommand{\arcsecpix}{arcsec\,pixel$^{-1}$}

\providecommand{\bjdtdb}{\ensuremath{\rm {BJD_{TDB}}}}

\providecommand{\msun}{\ensuremath{\mathrm M_{\sun}}}
\providecommand{\rsun}{\ensuremath{\mathrm R_{\sun}}}
\providecommand{\lsun}{\ensuremath{\mathrm L_{\sun}}}
\providecommand{\mj}{\ensuremath{\mathrm M_{\rm J}}}
\providecommand{\rj}{\ensuremath{\mathrm R_{\rm J}}}

\providecommand{\fave}{\langle F \rangle}
\providecommand{\fluxcgs}{10$^9$ erg s$^{-1}$ cm$^{-2}$}

\newcommand{\LSO}{La Silla Observatory}

\newcommand{\kepler}{{\it Kepler}}
\newcommand{\corot}{{\it CoRoT}}
\newcommand{\tess}{{\it TESS}}

\newcommand{\gaia}{{\it Gaia}}

\newcommand{\NGTS}{{\it NGTS}}

\newcommand{\exofast}{{\it EXOFASTv2}}
\newcommand{\emp}{\textsc{SpecMatch-emp}}

\usepackage{numprint}
\usepackage{graphicx}
\usepackage{adjustbox}
\usepackage{xcolor}
\usepackage{marvosym}
\usepackage{txfonts}

\begin{document}

   \title{Three short-period Jupiters from \tess}

  \subtitle{\Nplanets}

   \author{L. D. Nielsen\inst{1}\fnmsep\thanks{Louise.Nielsen@unige.ch}
        \and
        R.~Brahm \inst{2,3} 
        \and
        F.~Bouchy\inst{1}
        \and
        N.~Espinoza \inst{4} 
        \and
        O.~Turner\inst{1}
        \and
        S.~Rappaport \inst{5}
        \and
        L.~Pearce \inst{6} 
        \and 
        G.~Ricker \inst{5}
        \and
        R.~Vanderspek \inst{5}
        \and
        D.W.~Latham \inst{7}
        \and
        S.~Seager \inst{6,8,9}
        \and
        J.N.~Winn \inst{10}
        \and
        J.M.~Jenkins \inst{11}
        \and 
        J.S.~Acton \inst{12} 
        \and
       G. Bakos \inst{10} 
        \and
        T.~Barclay \inst{13,14} 
        \and
        K.~Barkaoui \inst{15,16} 
        \and
       W. Bhatti \inst{10} 
            \and
        C.~Brice\~{n}o \inst{17} 
        \and
        E.M.~Bryant \inst{18,19} 
        \and
        M.R.~Burleigh \inst{12} 
        \and 
        D.R.~Ciardi \inst{20} 
        \and 
        K.A.~Collins \inst{9} 
        \and
        K.I.~Collins \inst{21} 
        \and
        B.F.~Cooke \inst{18,19} 
        \and
       Z. Csubry \inst{10} 
       \and
        L.A.~dos~Santos \inst{1} 
        \and
        Ph. Eigm\"uller \inst{22} 
        \and  
        M.~M.~Fausnaugh \inst{5}
        \and
        T.~Gan \inst{23} 
        \and
        M.~Gillon \inst{15} 
        \and
        M.R.~Goad \inst{12} 
        \and
        N.~Guerrero \inst{5}
        \and 
        J.~Hagelberg \inst{1} 
       \and
        R.~Hart \inst{24} 
        \and
        T. Henning \inst{25} 
        \and
        C.X.~Huang \inst{5} 
        \and
        E.~Jehin \inst{26} 
         \and
        J.S.~Jenkins \inst{27,28} 
        \and 
        A.~Jord\'an \inst{2,3} 
        \and
        J.F.~Kielkopf \inst{29} 
        \and
       D. Kossakowski \inst{25} 
        \and
        B.~Lavie \inst{1} 
        \and
        N.~Law \inst{30} 
        \and
        M.~Lendl\inst{1,31}
        \and 
        J.P.~de~Leon \inst{32} 
        \and
        C.~Lovis \inst{1}
        \and
        A.W.~Mann \inst{30} 
        \and
        M.~Marmier \inst{1}
        \and
        J.~McCormac \inst{18,19} 
        \and
        M.~Mori \inst{32} 
        \and
         M.~Moyano \inst{33} 
         \and
        N.~Narita \inst{34,35,36,37,38} 
       \and
       D. Osip \inst{39} 
       \and
        J.F.~Otegi \inst{1,40}
       \and
        F.~Pepe \inst{1}
                \and
        F.J.~Pozuelos \inst{25,15} 
        \and
        L.~Raynard \inst{12} 
        \and
        H.M.~Relles \inst{7} 
         \and
       P.~Sarkis \inst{25} 
        \and
        D.~S{\'e}gransan \inst{1}
         \and
        J.V.~Seidel \inst{1} 
        \and
        A.~Shporer \inst{5} 
        \and 
        M.~Stalport \inst{1}
        \and 
        C.~Stockdale \inst{41} 
       \and
       V.~Suc \inst{2} 
        \and
        M.~Tamura \inst{32,34,36} 
        \and
        T.G.~Tan \inst{42} 
       \and
        R.H.~Tilbrook \inst{12} 
        \and
        E.B.~Ting \inst{11} 
        \and
        T.~Trifonov \inst{25} 
        \and
        S.~Udry \inst{1}
        \and
        A.~Vanderburg \inst{43} 
        \and
        P.J.~Wheatley \inst{18,19} 
        \and
        G.~Wingham \inst{44} 
        \and 
        Z.~Zhan \inst{8,9} 
        \and
        C.~Ziegler \inst{45} 
        }

   \institute{Geneva Observatory, University of Geneva, Chemin des Mailettes 51, 1290 Versoix, Switzerland
       \and 
    Facultad de Ingeniería y Ciencias, Universidad Adolfo Ib\'a\~nez, Av.\ Diagonal las Torres 2640, Pe\~nalol\'en, Santiago, Chile
   \and 
    Millennium Institute for Astrophysics, Chile
   \and 
   Space Telescope Science Institute, 3700 San Martin Drive, Baltimore, MD 21218, USA
   \and 
  Department of Physics and Kavli Institute for Astrophysics and Space Research, MIT, Cambridge, MA 02139, USA
   \and 
    NSF Graduate Research Fellow, Steward Observatory, University of Arizona, Tucson, AZ 85721, USA
     \and 
  Center for Astrophysics ${\rm \mid}$ Harvard {\rm \&} Smithsonian, 60 Garden Street, Cambridge, MA 02138, USA
  \and 
  Department of Earth, Atmospheric and Planetary Sciences, MIT, Cambridge, MA 02139, USA
      \and 
    Department of Aeronautics and Astronautics, MIT, Cambridge, MA 02139, USA
    \and 
  Department of Astrophysical Sciences, Princeton University, 4 Ivy Lane, Princeton, NJ 08544, USA
  \and 
   NASA Ames Research Center, Moffett Field, CA 94035, USA
        \and 
    Department of Physics and Astronomy, University of Leicester, University Road, Leicester, LE1 7RH, UK
        \and  
    NASA Goddard Space Flight Center, 8800 Greenbelt Rd, Greenbelt, MD 20771, USA
        \and 
    University of Maryland, Baltimore County, 1000 Hilltop Cir, Baltimore, MD 21250, USA
        \and 
    Astrobiology Research Unit, Université de Li\`{e}ge, 19C All\'{e}e du 6 Ao\^{u}t, 4000 Li\`{e}ge, Belgium
        \and 
    Oukaimeden Observatory, High Energy Physics and Astrophysics Laboratory, Cadi Ayyad University, Marrakech, Morocco   
       \and 
   Cerro Tololo Inter-American Observatory, Casilla 603, La Serena, Chile
   \and 
    Dept. of Physics, University of Warwick, Gibbet Hill Road, Coventry CV4 7AL, UK
    \and 
    Centre for Exoplanets and Habitability, University of Warwick, Gibbet Hill Road, Coventry CV4 7AL, UK
       \and 
    Caltech IPAC – NASA Exoplanet Science Institute 1200 E. California Ave, Pasadena, CA 91125, USA
       \and 
   George Mason University, 4400 University Drive, Fairfax, VA, 22030 USA
       \and 
    Institute of Planetary Research, German Aerospace Center, Rutherfordstrasse 2, 12489 Berlin, Germany
       \and 
    Physics Department and Tsinghua Centre for Astrophysics, Tsinghua University, Beijing 100084, China
      \and 
   \Cross Centre for Astrophysics, University of Southern Queensland, Toowoomba, QLD, 4350, Australia.
       \and 
    Max-Planck-Institut f\"ur Astronomie, K\"onigstuhl 17, Heidelberg 69117, Germany
        \and 
    Space Sciences, Technologies and Astrophysics Research (STAR) Institute, Universit\`{e} de Li\`{e}ge, 19C All\'{e}e du 6 Ao\^{u}t, 4000 Li\`{e}ge, Belgium 
    \and 
    Departamento de Astronomía, Universidad de Chile, Camino El Observatorio 1515, Las Condes, Santiago, Chile 
    \and 
    Centro de Astrof\'isica y Tecnolog\'ias Afines (CATA), Casilla 36-D, Santiago, Chile
    \and 
   Department of Physics and Astronomy, University of Louisville, Louisville, KY 40292, USA
   \and 
   Department of Physics and Astronomy, University of North Carolina at Chapel Hill, Chapel Hill, NC 27599-3255, USA
       \and 
    Space Research Institute, Austrian Academy of Sciences, Schmiedlstr. 6, 8042 Graz, Austria
       \and 
   The University of Tokyo, 7-3-1 Hongo, Bunkyō, Tokyo 113-8654, Japan
       \and 
    Instituto de Astronom\'ia, Universidad Cat\'olica del Norte, Angamos 0610, Antofagasta 1270709, Chile
   \and 
    Astrobiology Center, 2-21-1 Osawa, Mitaka, Tokyo 181-8588, Japan
    \and 
    JST, PRESTO, 2-21-1 Osawa, Mitaka, Tokyo 181-8588, Japan
    \and 
    National Astronomical Observatory of Japan, 2-21-1 Osawa, Mitaka, Tokyo 181-8588, Japan
    \and 
    Instituto de Astrofisica de Canarias (IAC), 38205 La Laguna, Tenerife, Spain
       \and 
   { Komaba Institute for Science, The University of Tokyo, 3-8-1 Komaba, Meguro, Tokyo 153-8902, Japan}
   \and 
    Las Campanas Observatory, Carnegie Institution of Washington, Colina el Pino, Casilla 601 La Serena, Chile
    \and 
    Institute for Computational Science, University of Zurich, Winterthurerstr. 190, CH-8057 Zurich, Switzerland
        \and 
    Hazelwood Observatory, Australia
        \and 
    Perth Exoplanet Survey Telescope, Perth, Western Australia
       \and 
    NASA Sagan Fellow, Department of Astronomy, Unversity of Texas at Austin, Austin, TX, USA
        \and 
    Mt. Stuart Observatory, New Zealand
   \and 
   Dunlap Institute for Astronomy and Astrophysics, University of Toronto, 50 St. George Street, Toronto, Ontario M5S 3H4, Canada
   }

   \date{Received 12 March 2020; accepted 28 May 2020}

  \abstract
  {We report the confirmation and mass determination of three hot Jupiters discovered by the Transiting Exoplanet Survey Satellite (\tess) mission:
  HIP~65Ab (TOI-129, TIC-201248411) is an ultra-short-period Jupiter orbiting a bright ($V$=11.1 mag) K4-dwarf every 0.98 days. It is a massive $3.213 \pm 0.078\ \mj$ planet in a grazing transit configuration with an impact parameter of $b=1.17^{+0.10}_{-0.08}$. As a result the radius is poorly constrained, $2.03 ^{+0.61}_{-0.49}\ \rj$.
  The planet's distance to its host star is less than twice the separation at which it would be destroyed by Roche lobe overflow. It is expected to spiral into HIP~65A on a timescale ranging from 80 Myr to a few gigayears, assuming a reduced tidal dissipation quality factor of $Q^{\prime}_s = 10^7 - 10^9$.
  We performed a full phase-curve analysis of the \tess\ data and detected both illumination- and ellipsoidal variations as well as Doppler boosting. HIP~65A is part of a binary stellar system, with HIP~65B separated by 269 AU (3.95 arcsec on sky).
  TOI-157b (TIC 140691463) is a typical hot Jupiter with a mass of $1.18 \pm 0.13\ \mj$ and a radius of $1.29 \pm 0.02\ \rj$. It has a period of 2.08 days, which corresponds to a separation of just 0.03 AU. This makes TOI-157 an interesting system, as the host star is an evolved G9 sub-giant star (V=12.7).
  TOI-169b (TIC 183120439) is a bloated Jupiter orbiting a V=12.4 G-type star. It has a mass of $0.79 \pm 0.06\ \mj$ and a radius of $1.09 ^{+0.08}_{-0.05}\ \rj$. Despite having the longest orbital period ($P=2.26$~days) of the three planets, TOI-169b receives the most irradiation and is situated on the edge of the Neptune desert. All three host stars are metal rich with \feh\ ranging from 0.18 - 0.24.}

   \keywords{Planets and satellites: detection --
   Planets and satellites: individual: (TOI-129, TIC 201248411, HIP 65A),
   Planets and satellites: detection --
   Planets and satellites: individual: (TOI-157, TIC 140691463),
   Planets and satellites: detection --
   Planets and satellites: individual: (TOI-169, TIC 183120439)}

   \maketitle
%

\section{Introduction}
Since July 2018, the Transiting Exoplanet Survey Satellite \citep[\tess\ -][]{Ricker:2015} has surveyed the Southern and Northern Hemispheres for exoplanets transiting bright stars. Based on the first year of observations in the south (Sectors 1 - 13), a total number of 1117 \tess\ Objects-of-Interest \citep[TOIs,][]{TOIcatalog} have been identified. Currently, 667 of these are still considered as planet candidates and 55 have been confirmed as new \tess-planets and four as transiting brown dwarfs \citep[including studies in preparation and published results, see e.g.][]{2019ApJ...877L..29C,2019A&A...625A..16J,2019A&A...623A.165E,2019NatAs...3.1099G,2020MNRAS.tmp..148E,2020MNRAS.tmp..252D,nielsen2020,subjak2019,2020arXiv200201943C}. We note that 146 of the TOIs from Sectors 1 - 13 are previously known planets. 

A recent study by \cite{Zhou2019} offers a first estimate of the occurrence rate of hot Jupiters discovered by \tess\ by analysing a sample of bright ($T_{\rm mag} <10$) main sequence stars observed by \tess. They find an occurrence rate of $0.40 \pm 0.10 \%$ which is in agreement with statistics based on the \kepler\ mission \citep{Fressin2013, 2016A&A...587A..64S}. An even rarer sub-population of hot Jupiters are the ultra-short-period (USP) Jupiters with orbital periods shorter than one day. To date, the following eight such planets are known: WASP-18b \citep{wasp18}, WASP-19b \citep{wasp19}, WASP-43b \citep{wasp43}, WASP-103b \citep{wasp103}, HATS-18b \citep{hats18}, KELT-16b \citep{kelt16}, NGTS-6b \citep{ngts6}, and NGTS-10b \citep{ngts10}. 

Hot Jupiters, and in particular USP Jupiters, can offer insights into planet-star interactions such as photo-evaporation and atmospheric escape \citep{2020MNRAS.tmp..240B,2018MNRAS.479.5012O,2009ApJ...693...23M}, atmospheric structure and chemistry \citep{2018A&A...617A.110P,2015ApJ...801...86K,Kreidberg14,2014A&A...563A..41M}, and tidal decay \citep{yee2020}. These planets shape the upper edge of the Neptune desert \citep{Mazeh2016,Szabo2011} which constitutes a dearth of sub-Jovian planets at short periods. The proposed mechanisms creating the desert are numerous, but they can generally be regarded as a combination of the following three dominant processes: photo-evaporation which strips less massive planets of their outer layers \citep{2016NatCo...711201L,2018MNRAS.479.5012O}; the availability of disc material during planet formation \citep{Armitage2007}; and planet migration \citep{Demangeon2018,2009ApJ...704..989A}.

Massive close-in planets also challenge current planet formation models; they represent the bulk of the mass and angular momentum in their systems while shaping their formation and evolution over time. An extreme case is NGTS~1b, which is a hot Jupiter around a M0 star \citep{ngts1}. Both in-situ formation and scenarios where the Jupiter is formed far out in the system followed by subsequent inward migration are still being considered in order to explain the presence of hot Jupiters \citep{2018ApJ...866L...2B,2008ApJ...678..498N}.

In this work, we present one USP and two hot Jupiters orbiting bright stars observed by \tess\ in its first year of operation. Table \ref{tab:stellar} lists the host stars stellar parameters. We modelled the systems self consistently with \exofast\ using transit light curves and radial velocity (RV) measurements to obtain masses and radii for all three systems. Our analysis is based on RV data from the high resolution spectrographs CORALIE on the Swiss 1.2 m telescope and FEROS on the 2.2 m MPG/ESO telescope, both in La Silla, Chile. In addition to the \tess\ data, we also utilise data from ground based photometric facilities that are part of the TESS Follow-up Observing Program, including LCOGT, NGTS, CHAT, Trappist-South, IRSF, PEST, Mt. Stuart Observatory, MKO, and Hazelwood Observatory. Additionally, SOAR speckle imaging was used to rule out close stellar companions.

\begin{table*}
\centering
\caption{\label{tab:stellar} Stellar Properties for \Nstars. Results for stellar parameters modelled in this study can be found in Table \ref{tab:results}.}
	\begin{tabular}{lcccc}
	\hline\hline 
	Property	& HIP~65A	 & TOI-157 &	TOI-169 &	Source\\
	\hline
	Spectral type & K4V & G9IV & G1V & \cite{Pecaut:2013}\\
    2MASS ID & J00004490-5449498            & J04544830-7640498     & J01070679-7511559 & 2MASS\\
    \gaia\ ID DR2 & 4923860051276772608     & 4624979393181971328   &	4684513614202233728        & \gaia \\
    TIC  ID & 201248411                     & 140691463             &  183120439        & \tess \\
    TOI &  TOI-129                          & TOI-157               & TOI-169   &\tess \\
    \\
    \multicolumn{3}{l}{Astrometric Properties}\\
    R.A.		&	00:00:44.56  		&  04:54:48.34   & 01:07:06.88   & \tess	\\
	Dec.			&	-54:49:50.93	    & -76:40:50.17   & -75:11:56.19  & \tess	\\
    $\mu_{{\rm R.A.}}$ (\masy)&    $ -202.82 \pm 0.03 $ &  $11.71 \pm 0.035$ & $ 20.23 \pm 0.05 $  & \gaia \\
	$\mu_{{\rm Dec.}}$ (\masy)&    $ -71.52 \pm 0.03 $ &  $-18.92 \pm 0.05$ & $ -15.61\pm 0.05 $  & \gaia\\
    Parallax  (mas) &  $16.156 \pm 0.021$  &  $2.783 \pm 0.022$ & $2.424\pm0.025$   & \gaia\\
    \\
    \multicolumn{3}{l}{Photometric Properties}\\
	B (mag)				&$ 12.29 \pm 0.15$  & $ 13.45 \pm 0.03 ^{\dagger}$ & $ 13.06 \pm 0.02 $ 	& Tycho / APASS$^{\dagger}$ \\
	V (mag)				&$ 11.13 \pm 0.06 $ & $ 12.73 \pm 0.05 ^{\dagger}$ & $ 12.36 \pm 0.05 $ 	& Tycho / APASS$^{\dagger}$\\ 
    G (mag)				&$10.590\pm 0.004 $ & $12.514 \pm 0.003 $ & $ 12.238 \pm 0.018 $ 	& \gaia\\
    T (mag)	    		&$ 9.901\pm 0.006 $  & $12.017 \pm 0.006 $ & $ 11.788 \pm 0.006 $ 	& \tess\\
    J (mag)				&$ 8.92 \pm 0.02 $  & $ 11.37 \pm 0.02 $ & $ 11.18 \pm 0.01 $ 	& 2MASS\\
   	H (mag)				&$ 8.40 \pm 0.03 $  & $ 10.99 \pm 0.03 $ & $ 10.90 \pm 0.02 $ 	& 2MASS\\
	K$_{\rm s}$ (mag) 	&$ 8.29 \pm 0.03 $  & $ 10.89 \pm 0.02 $ & $ 10.82 \pm 0.02 $ 	& 2MASS\\
    W1 (mag)			&$ 8.12 \pm 0.02 $  & $ 10.85 \pm 0.03 $ & $ 10.80 \pm 0.03 $ 	& WISE\\
    W2 (mag)			&$ 8.18 \pm 0.02 $  & $ 10.89 \pm 0.03 $ & $ 10.84 \pm 0.03 $ 	& WISE\\
    W3 (mag)    		&$ 8.10 \pm 0.02 $  & $ 10.87 \pm 0.07 $ & $ 10.75 \pm 0.08 $  	& WISE\\
    W4 (mag)    		&$ 8.12 \pm 0.21 $  &                      & 	& WISE \\
    \\
	\hline
    \end{tabular}
\tablefoot{Tycho \citep{Tycho}; 2MASS \citep{2MASS}; WISE \citep{WISE}; \gaia\ \citep{Gaia2018}; APASS \citep{apass}. Spectral type is based on \teff\ from global modelling (see Section \ref{sec:exofast}) and Table 5 in \cite{Pecaut:2013}.}
\end{table*}


\section{Observations}
A summary of all the data used in the joint analysis of \Nplanets\ can be found in Table \ref{tab:obs}. Additionally SOAR speckle imaging was used to rule out close stellar companions, as described in Sect. \ref{sec:speckle}.

\begin{table}
\caption{\label{tab:obs} Summary of the discovery \tess-photometry, follow-up photometry and radial velocity observations of \Nstars. }

\begin{tabular}{lcr}\hline
Date  & Source & N.Obs / Filter\\
\hline 
{\bf HIP~65A} (TOI-129) & & \\
2018 July -- Sep & \tess\ 2 min& TESS  \\
2018 Nov -- Dec & CORALIE & 17  \\ 
2018 Nov -- Dec & FEROS &  17     \\ 
2018 Sep 7 & LCO-SSO & $ z\prime$ \\
2018 Sep 10 & MKO & $r\prime$ \\
2018 Sep 13 & LCO-SSO & B \\
2018 Sep 13 & LCO-SSO & $i \prime$ \\
2018 Sep 14 & LCO-SSO & B \\
2018 Sep 14 & LCO-SSO & $i \prime$ \\
2018 Sep 14 & PEST & V \\
2018 Nov 30 & NGTS & NGTS \\
2018 Dec 2 & NGTS & NGTS \\

\hline 
\multicolumn{1}{l}{\bf TOI-157} & \\
2018 July -- 2019 Feb & \tess\ FFI & TESS \\
2019 Mar -- July & \tess\ 2 min & TESS  \\
2018 Nov -- 2019 Jan & CORALIE & 24  \\ 
2018 Nov -- Dec  & FEROS &  2  \\ 
2018 Sep 15 & LCO-SAAO & $i\prime$ \\
2018 Sep 21 & LCO-CTIO 0.4 m & $i\prime$ \\
2018 Oct 07 & IRSF & H \\
2018 Oct 07 & IRSF & J \\
2018 Oct 18 & MtStuart & $g\prime$ \\
2018 Oct 22 & Hazelwood & Ic \\
2018 Oct 24 & Hazelwood & Ic \\
2018 Nov 08 & LCO-CTIO & $g\prime$ \\
2018 Nov 08 & LCO-CTIO & $i\prime$ \\

\hline 
\multicolumn{1}{l}{\bf TOI-169} & \\
2018 July 25 - Sep 20 & \tess\ FFI & TESS  \\
2019 Jun -- Jul & \tess\ 2 min & TESS \\
2018 Oct -- Nov  & FEROS &  10  \\
2019 Jun -- Jul  & CORALIE & 6  \\
2018 Sep 11  & LCO-SAAO  & $i\prime$ \\
2018 Sep 26  & LCO-SAAO  & $i\prime$ \\
2018 Oct 01  & CHAT  & $i\prime$ \\
2018 Nov 03  & Trappist-South    & B \\
2018 Nov 13  & LCO-CTIO  & $g\prime$ \\
\hline 

\end{tabular}
\end{table}

\subsection{Discovery photometry from TESS}
\Nstars\ were all observed by \tess\ in multiple Sectors and announced as TOIs from Sector 1 by the \tess\ Science Office. 
HIP~65A (TOI-129, TIC~201248411) was observed with 2-min cadence in Sectors 1 and 2 from 2018-Jul-25 to 2018-Sep-20. TOI-157 (TIC 140691463) was observed in Sectors 1-8 in the full frame images (FFI) with 30-min cadence and in Sectors 9, 11, 12, and 13 with 2-min cadence. TOI-169 was observed in the FFIs in Sector 1 and later in Sector 13 with 2-min cadence. 

For the Sectors with 2-min data available we use the publicly available Simple Aperture Photometry flux with Pre-search Data Conditioning (PDC-SAP) \citep{Stumpe:2014,Stumpe:2012,2012PASP..124.1000S,Jenkins:2010} provided by the Science Processing Operations Center \citep[SPOC - ][]{Jenkins:2016}. 
For the FFI data we utilised light curves produced by the MIT Quick Look pipe-line .

 
 

\subsection{Follow-up spectroscopy with CORALIE \& FEROS}
\Nstars\ were observed with the high resolution spectrograph CORALIE on the Swiss $1.2\, \mathrm{m}$ Euler telescope at \LSO~\citep{CORALIE}. CORALIE is fed by a 2\arcsec\ fibre and has a resolution of $R=60,000$. 
RVs and line bisector spans were calculated via cross-correlation with a G2 binary mask, using the standard CORALIE data-reduction pipeline. 

The three systems were also monitored with the FEROS spectrograph \citep{feros} mounted on the MPG 2.2m telescope installed at \LSO. FEROS has a spectral resolution of $R=48,000$ and is fibre fed from the telescope. Observations were performed with the simultaneous calibration mode where a second fibre is illuminated by a Thorium-Argon lamp in order to trace the instrumental RV drift. 17, 2, and 10 FEROS spectra were obtained for \Nstars, respectively. FEROS data were processed with the CERES pipeline \citep{ceres}, which delivers precision RVs computed via the cross-correlation technique.

\begin{figure} 
   \centering   
  \includegraphics[width=\columnwidth,trim={0cm 0cm 1cm 13cm},clip]{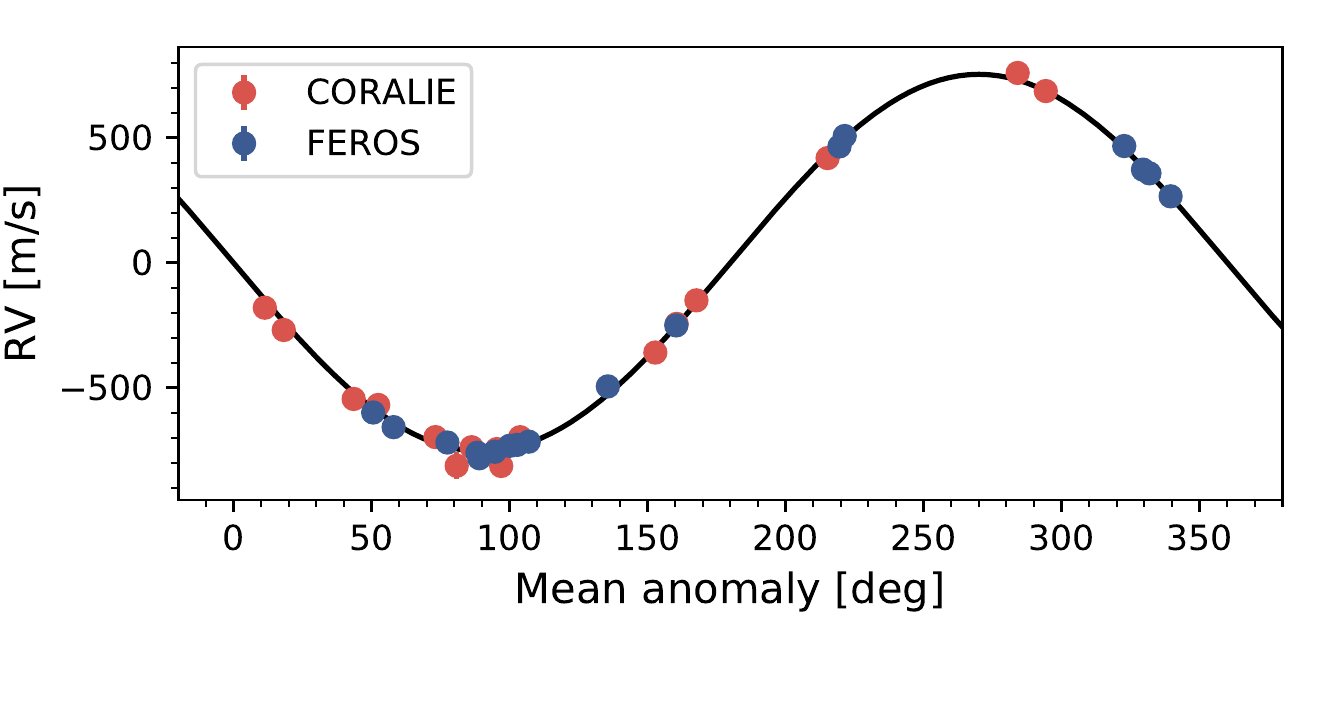}
      \caption{\label{fig:129_RV} RVs from CORALIE and FEROS for HIP~65A, phase folded on the ephemeris of the planet. Error bars are included, but too small to show.}
\end{figure}

The first few RV measurements were used for reconnaissance, to check for a visual or spectroscopic binary. Once a significant change in RV had been identified to be  consistent with the ephemerides provided by \tess\, we commenced intensive follow-up observations. 
The RVs from both CORALIE and FEROS are listed in Appendix~\ref{sec:RVdata} (online version only). One CORALIE measurement from BJD 58460.665537 (\numprint{-2400000}) was excluded from the global analysis due to low signal-to-noise ratio (S/N). In Figs. \ref{fig:129_RV}, \ref{fig:157_RV} and \ref{fig:169_RV} we plot the phase folded RVs along with our best-fit model from the joint analysis (Sect. \ref{sec:exofast}).

The Lomb-Scargle periodograms for the RV measurements show significant signals (above 0.1 \% False Alarm Probability, FAP) at the orbital periods recovered from the transit data for all three systems. To ensure that the RV signal does not originate from cool stellar spots or a blended eclipsing binary, we checked for correlations between the line bisector span and the RV measurements \citep{Queloz2001}. We found no evidence for correlation for any of our targets. None of the Lomb-Scargle periodograms for the activity indicators have peaks above 10\% FAP.

\subsection{Follow-up photometry}

We acquired ground-based time-series follow-up photometry of \Nstars\ as part of the TESS Follow-up Observing Program (TFOP) to attempt to (1) rule out nearby eclipsing binaries (NEBs) as potential sources of the TESS detection, (2) detect the transit-like event on target to confirm the event depth and thus the TESS photometric deblending factor, (3) refine the TESS ephemerides, (4) provide additional epochs of transit centre time measurements to supplement the transit timing variation (TTV) analysis, and (5) place constraints on transit depth differences across optical filter bands. We used the {\tt TESS Transit Finder}, which is a customised version of the {\tt Tapir} software package \citep{Jensen:2013}, to schedule our transit observations. 

\subsubsection{Las Cumbres Observatory Global Telescope (LCOGT)}

Five, four, and three full transits of \Nstars, respectively, were observed using the Las Cumbres Observatory Global Telescope (LCOGT) 1.0-m and 0.4-m network \citep{Brown:2013} nodes at Cerro Tololo Inter-American Observatory (CTIO), Siding Spring Observatory (SSO), and South Africa Astronomical Observatory (SAAO). The 1.0-m telescopes are equipped with $4096\times4096$ LCO SINISTRO cameras having an image scale of 0$\farcs$389 pixel$^{-1}$ resulting in a $26\arcmin\times26\arcmin$ field of view. The 0.4-m telescopes are equipped with $2048\times3072$ SBIG STX6303 cameras having an image scale of 0$\farcs$57 pixel$^{-1}$ resulting in a $19\arcmin\times29\arcmin$ field of view. The images were calibrated using the standard LCOGT BANZAI pipeline \citep{McCully:2018}. The photometric data were extracted using the {\tt AstroImageJ} ({\tt AIJ}) software package \citep{Collins:2017}.

HIP~65A was observed five times using the SSO 1.0-m network node on 2018-Sep-7 in Pan-STARSS $z$-short band, 2018-Sep-13 in B-band and $i'$-band, and 2018-Sep-14 in B-band and $i'$-band. The HIP~65Ab transit was detected on-target using photometric apertures with radius as small as 1$\farcs$2. Since the typical stellar FWHM in the images is 2$\farcs$1, most of the flux from the closest \gaia\ DR2 neighbour 3$\farcs$95 to the north-west, which is 4.4 magnitudes fainter in TESS band, was excluded from the follow-up target star aperture. Thus, all known neighbouring \gaia\ DR2 stars are ruled out as the source of the on-target transit detection.

TOI-157 was observed using the SAAO 1.0-m network node on 2018-Sep-15 in $i'$-band, two times using the CTIO 1.0-m network node on 2018-Sep-8 in $g'$-band and $i'$-band, and one time using the CTIO 0.4-m network node on 2018-Sep-21 in $i'$-band. The TOI-157 transit was detected on-target using photometric apertures with radius as small as 4$\farcs$7, which rules out all known neighbouring \gaia\ DR2 stars as the source of the transit detection.

TOI-169 was observed using the SAAO 1.0-m network node on 2018-Sep-11 in $i'$-band, the CTIO 1.0-m network node on 2018-Sep-13 in $g'$-band, and the SAAO 0.4-m network node on 2018-Sep-26 in $i'$-band. The TOI-169 transit was detected on-target using photometric apertures with radius as small as 2$\farcs$7, which rules out all known neighbouring \gaia\ DR2 stars as the source of the transit detection.

\subsubsection{Next Generation Transit Survey (NGTS)}
Two full transits of HIP~65Ab were observed using the Next Generation Transit Survey \citep[\NGTS,][]{Wheatley2018} on the nights UT 2018-Nov-30 and 2018-Dec-02. On both nights, a single 0.2~m NGTS telescope was used. Across the two nights, a total of 2422 images were obtained using the custom NGTS filter (520 - 890 nm) and an exposure time of 10 seconds. We had sub-pixel level stability of the target on the CCD, thanks to the telescope guiding performed by the DONUTS algorithm \citep{McCormac2013}. The data reduction was performed using a custom aperture photometry pipeline. For the reduction, comparison stars, which were similar to HIP~65A in both apparent magnitude and colour, were automatically selected.

\subsubsection{Chilean-Hungarian Automated Telescope (CHAT)}
A full transit of TOI-169 was obtained with the 0.7 m Chilean-Hungarian Automated Telescope (CHAT) installed at Las Campanas Observatory in Chile.
The observations took place on the night of 2018-Oct-01, using the sloan $i$ filter and an exposure time of 130 s. The 60 science images where processed with a dedicated pipeline which is an adaptation of the routines developed for the processing of photometric time series with LCOGT facilities \citep[see][]{hartman:2019,jordan:2019,espinoza:2019}. This pipeline automatically determines the optimal aperture for the photometry, which was 7 pixels in this case (4$\farcs$2). The obtained per point precision was 1100 ppm, which was enough to detect the $\approx$6 mmag transit, confirming that this was the source of the signal detected by TESS. 

\subsubsection{TRAPPIST-South}
TRAPPIST-South at ESO-La Silla Observatory in Chile is a 60~cm Ritchey-Chretien telescope, which has a thermoelectrically cooled $2K\times2K$ FLI Proline CCD camera with a field of view of $22'\times22'$ and pixel-scale of 0.65 \arcsecpix \citep[for more detail, see][]{jehin2011,gillon2013}. We carried out a full-transit observation of TOI-169 on 2018-Nov-03 with $B$ filter with an exposure time of 50~s. We took 220 images and made use of {\tt AIJ} to perform aperture photometry. The optimum aperture being 7~pixels (4$\farcs$55) and a PSF of 2$\farcs$80. We confirmed the event on the target star and we cleared all the stars of eclipsing binaries within 2.5~arcmin around the target star.  

\subsubsection{Infrared Survey Facility (IRSF)}
TOI-157 was observed with the Infrared Survey Facility (IRSF) 1.4 m telescope located in Sutherland, South Africa on UT 2018-Oct-7.
We used the Simultaneous Infrared Imager for Unbiased Survey (SIRIUS: \citealt{2003SPIE.4841..459N}) camera for the observation, which is equipped with two dichroic mirrors and can take $J$, $H$, and $K_{\rm s}$ bands simultaneously with three 1K$\times$1K HgCdTe detectors. On the observing night, the $K_{\rm s}$ band detector had a trouble, and only $J$ and $H$ band data were useful.
We took 300 frames for each band with an exposure time of 60 seconds. We used a position locking software introduced in \citet{2013PASJ...65...27N} during the observation. We applied a dedicated pipeline for the SIRIUS data\footnote{http://irsf-software.appspot.com/yas/nakajima/sirius.html} to make sky flats.
Dark subtraction, flat fielding, and subsequent standard aperture photometry were done with a customised pipeline by \citet{2011PASJ...63..287F}.

\subsubsection{Perth Exoplanet Survey Telescope (PEST)}

We observed a full transit of HIP~65Ab on UTC 2018-Sep-14 in V-band from the Perth Exoplanet Survey Telescope (PEST) near Perth, Australia. The 0.3 m telescope is equipped with a $1530\times1020$ SBIG ST-8XME camera with an image scale of 1$\farcs$2 pixel$^{-1}$ resulting in a $31\arcmin\times21\arcmin$ field of view. A custom pipeline based on {\tt C-Munipack}\footnote{http://c-munipack.sourceforge.net} was used to calibrate the images and extract the differential photometry, using an aperture with radius 10$\farcs$6. The images have typical stellar point spread functions (PSFs) with a FWHM of $\sim4\arcsec$.  

\subsubsection{Mt. Stuart Observatory}

We observed a full transit of TOI-157 on UTC 2018-Oct-18 in $g'$-band from Mt. Stuart near Dunedin, New Zealand. The 0.32 m telescope is equipped with a $3072\times2048$ SBIG STXL6303E camera with an image scale of 0$\farcs$88 pixel$^{-1}$ resulting in a $44\arcmin\times30\arcmin$ field of view. {\tt AIJ} was used to calibrate the images and extract the differential photometry with an 8$\farcs$8 aperture radius. The images have typical stellar PSFs with a FWHM of $\sim5\arcsec$.

\subsubsection{Mt. Kent Observatory (MKO)}

We observed a full transit of HIP~65Ab on UTC 2018-Sep-10 in $r'$-band from Mt. Kent Observatory (MKO) near Toowoomba, Australia. The 0.7-m telescope is equipped with a $4096\times4096$ Apogee Alta F16 camera with an image scale of 0$\farcs$41 pixel$^{-1}$ resulting in a $27\arcmin\times27\arcmin$ field of view. {\tt AIJ} was used to calibrate the images and extract the differential photometry with a 3$\farcs$3 aperture radius. The images have typical stellar PSFs with a FWHM of $\sim2\arcsec$.

\subsubsection{Hazelwood Observatory}

Hazelwood Observatory is a backyard observatory located in Victoria, Australia. Photometric follow-up data for TOI-157 was obtained on 2018-Oct-22 and 24 in the Ic band, using a 0.32-m Planewave CDK telescope and SBIG STT3200 CCD camera, with 2148 x 1472 pixels (FoV 20\arcmin x 13\arcmin). The observations on 2018-Oct-22 covered a full transit with some observations missing near ingress and at mid-transit due to passing cirrus cloud. The observations taken on 2018-Oct-24 were not continuous due to passing cirrus cloud. The frames were corrected for Bias, Dark and Flat Fields using MaximDL. Differential photometry was extracted using {\tt AIJ}.

\begin{figure} 
   \centering   
  \includegraphics[width=\columnwidth,trim={0cm 0cm 1cm 13cm},clip]{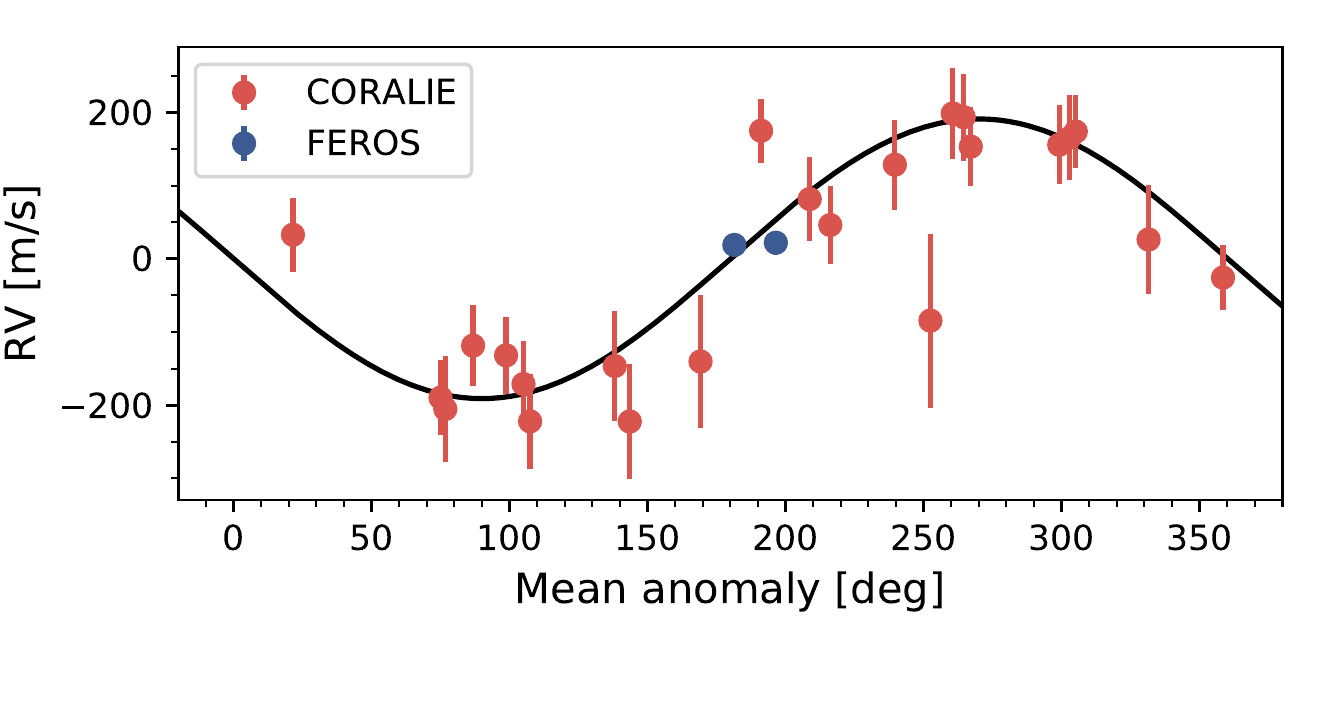}
      \caption{\label{fig:157_RV} CORALIE and FEROS RVs for TOI-157, phase folded on the ephemeris for TOI-157b.}
\end{figure}

\begin{figure} 
   \centering   
  \includegraphics[width=\columnwidth,trim={0cm 0cm 1cm 13cm},clip]{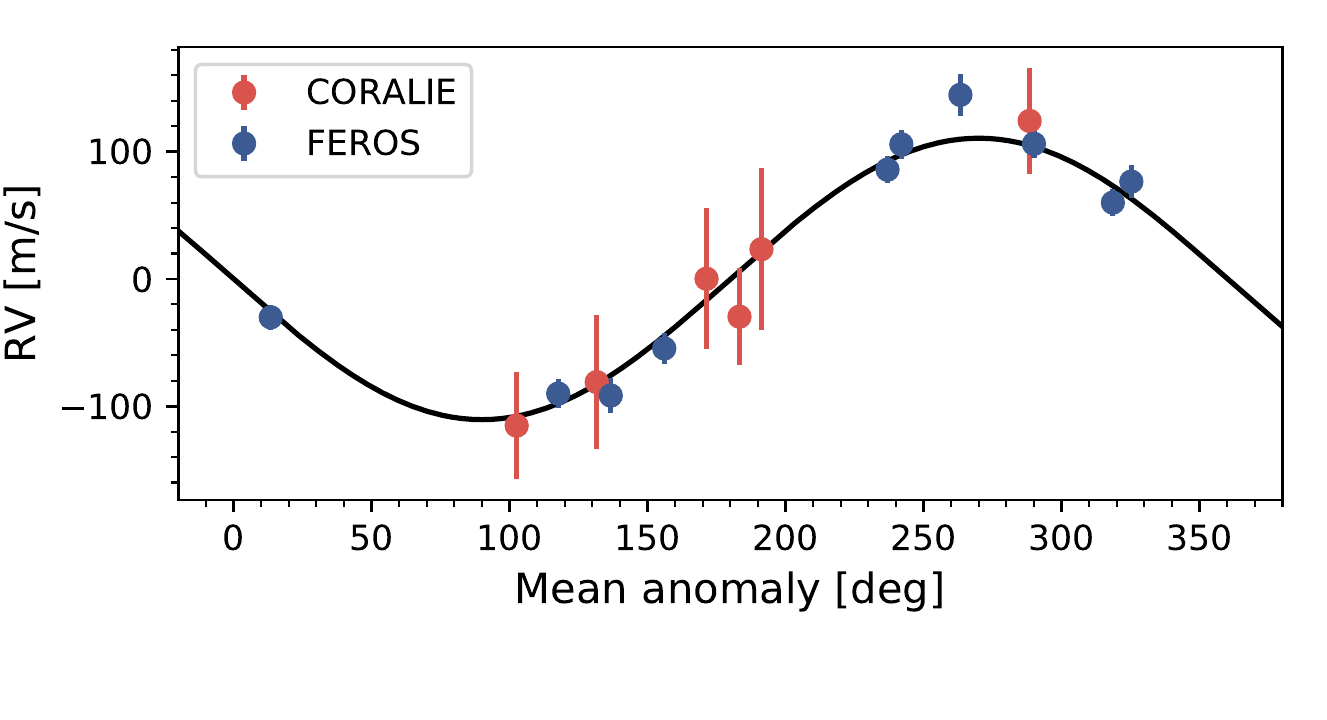}
      \caption{\label{fig:169_RV} CORALIE and FEROS RVs for TOI-169, phase folded on the ephemeris for TOI-169b.}
\end{figure}

\subsection{SOAR speckle imaging} \label{sec:speckle}

\begin{figure} 
   \centering   
  \includegraphics[width=0.95\columnwidth,trim={0cm 0cm 0cm 0cm},clip]{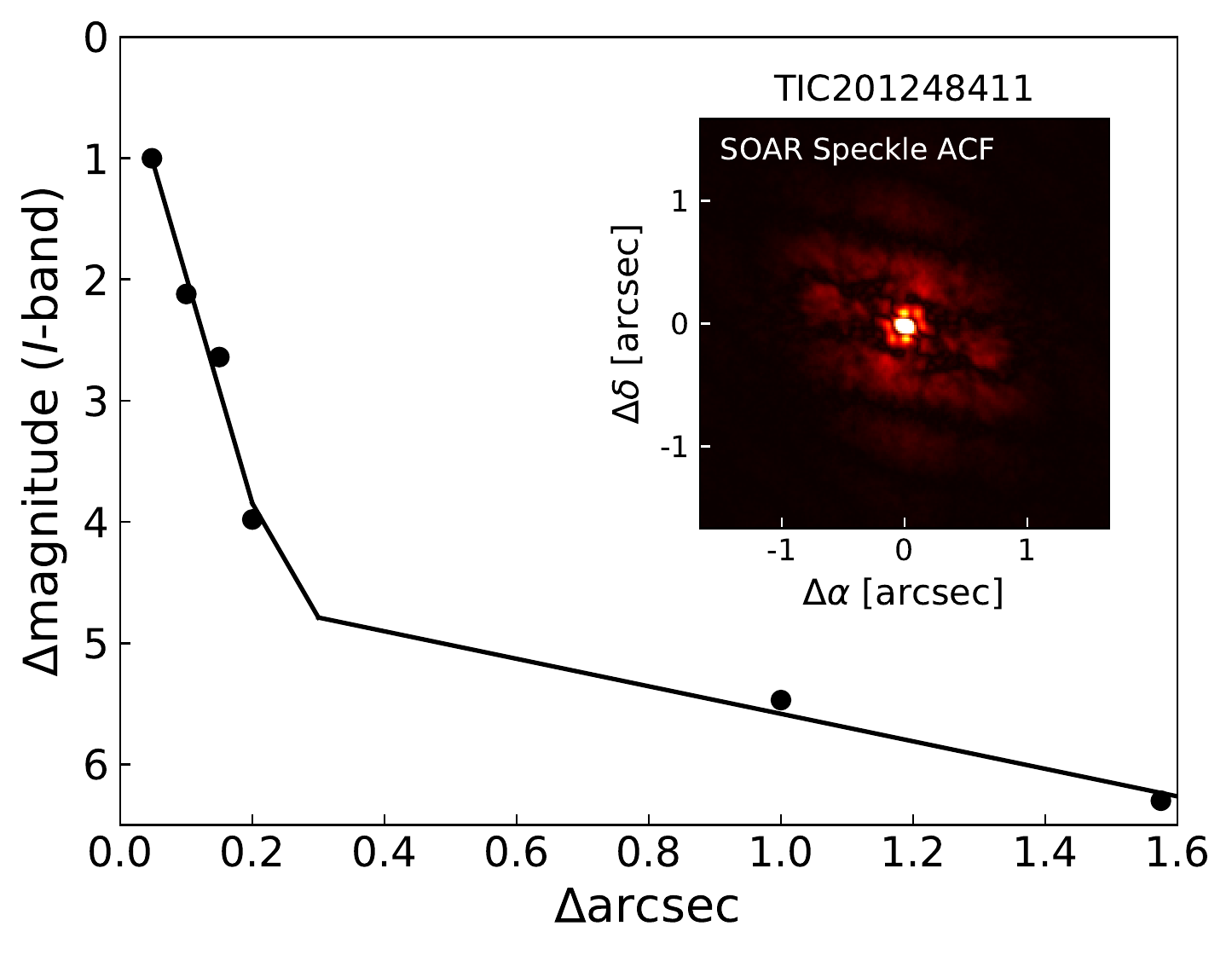}
  \includegraphics[width=0.95\columnwidth,trim={0cm 0cm 0cm 0cm},clip]{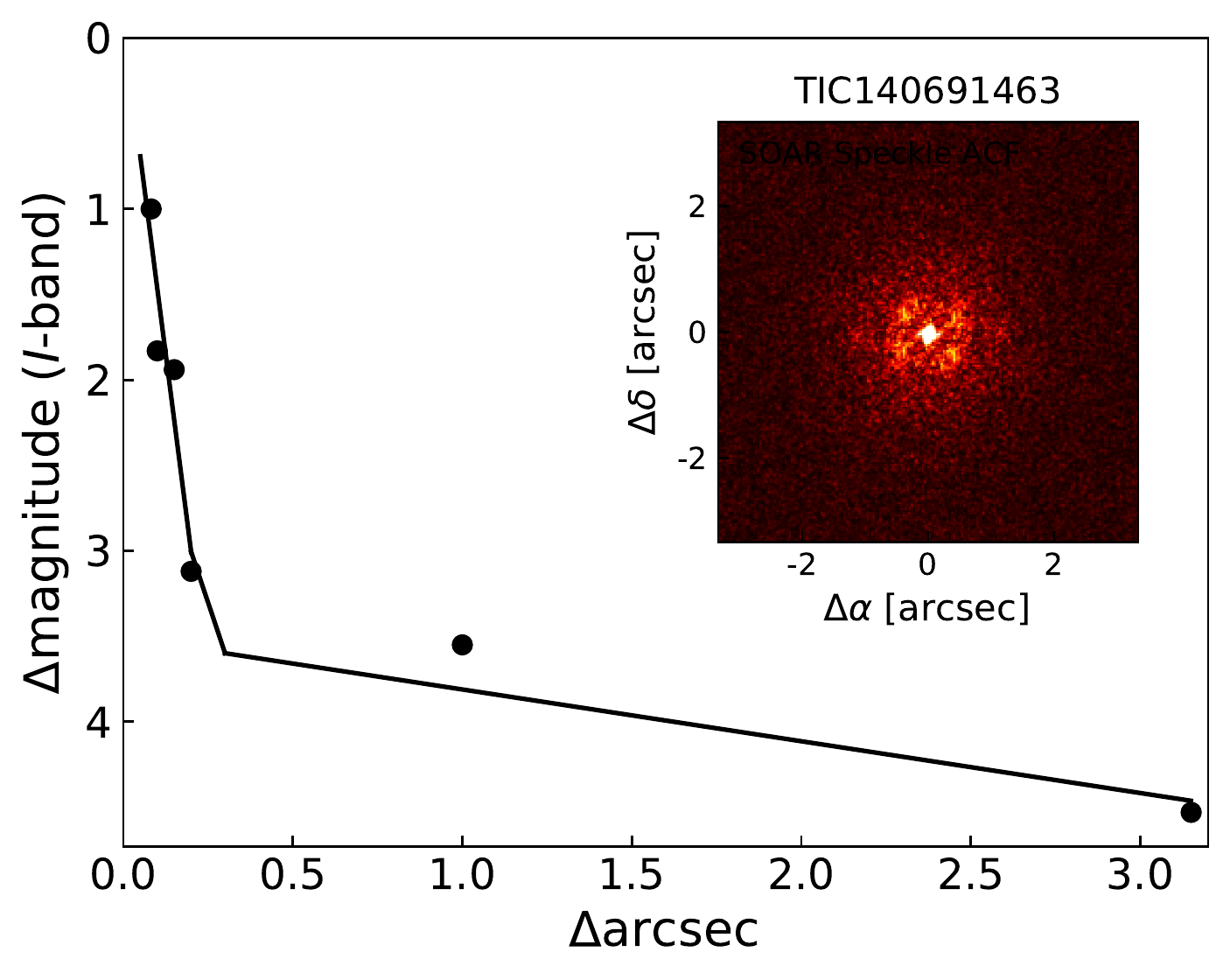}
  \includegraphics[width=0.95\columnwidth,trim={0cm 0cm 0cm 0cm},clip]{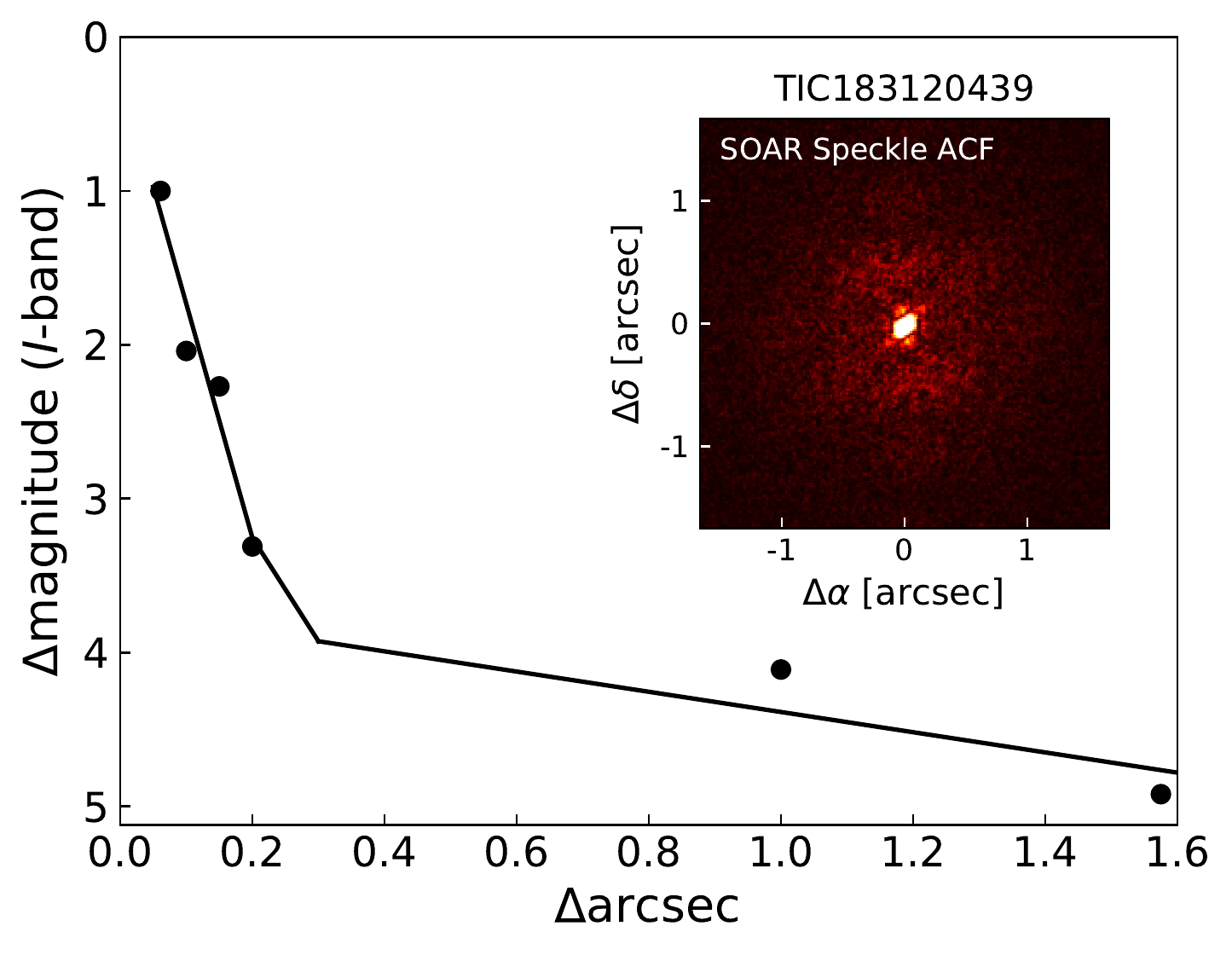}

      \caption{SOAR speckle imaging of \Nstars\ from the top.}
         \label{fig:speckle}
\end{figure}

TESS is in-sensitive to close companions due to its relatively large 21\arcsec pixels. Companion stars can contaminate the photometry, resulting in an underestimated planetary radius or may be the source of an astrophysical false positive. We searched for previously unknown companions to \Nstars\ with SOAR speckle imaging \citep{2018PASP..130c5002T} on UT 2018-Sep-25 and UT 2018-Oct-21, observing in a similar visible bandpass as \tess. Further details of the observations are available in \cite{2020AJ....159...19Z}.  We did not detect any nearby stars to the three host stars within 3\arcsec. The 5$\sigma$ detection sensitivity and the speckle auto-correlation function from the SOAR observations are plotted in Fig.~ \ref{fig:speckle}.

\section{Spectral analysis} \label{sec:specMatch}
Stellar atmospheric parameters, including effective temperature, \teff, surface gravity, $\log g$, and metalicity, [Fe/H], were derived using \emp\ \citep{specmatch} on stacked FEROS spectra for HIP~65A and TOI-169. For TOI-157, we ran \emp\ on stacked CORALIE spectra. 

\emp\ matches the input spectra to a vast library of stars with well-determined parameters derived with a variety of independent methods, such as interferometry, optical and NIR photometry, asteroseismology, and LTE analysis of high-resolution optical spectra. We used the spectral region around the Mg I b triplet (5100 - 5340 {\AA}) to match our spectrum to the library spectra through $\chi^2$ minimisation. A weighted linear combination of the five best matching spectra were used to extract \teff, $R_{\mathrm{s}}$ and \feh.

The projected rotational velocity of the star, $v \sin i$, was computed using the calibration between $v \sin i$ and the width of the CORALIE CCF from \cite{Santos2002}. The formal result was smaller than what can be resolved by CORALIE, and we can therefore only establish an upper limit of 2.5\, \kms. An independent analysis performed on the FEROS spectra using the CERES pipeline yields similar $v \sin i$ upper limits.

Chromospheric activity indicators $\log R^{\prime}_\mathrm{HK}$ were computed for each of the three stars using the FEROS spectra using the prescription in \citep{Boisse2009}. The average values are listed in Table \ref{tab:results}.












\begin{table*}
\begin{minipage}{\textwidth}
\caption{ \label{tab:results} Median values and 68\% confidence intervals of the posterior distributions from joint modelling for \Nstars\ as described in Sec. \ref{sec:exofast}. $P_{\mathrm{rot}}$, $v \sin i$ and $\log R^{\prime}_\mathrm{HK}$ are results from separate analyses, see Sec. \ref{sec:specMatch} and \ref{sec:HIP65A_rot}.}
\begin{adjustbox}{max width=\textwidth}
\begin{tabular}{lccccccccccccc} \hline \hline
\smallskip\\\multicolumn{2}{l}{Stellar Parameters:}& HIP~65A & TOI-157 & TOI-169\smallskip\\
~~~~$M_*$\dotfill &Mass (\msun)\dotfill  & $0.781\pm0.027$ & $0.948^{+0.023}_{-0.018}$ & $1.147^{+0.069}_{-0.075}$ \\
~~~~$R_*$\dotfill &Radius (\rsun)\dotfill  & $0.7242^{+0.0081}_{-0.0091}$ & $1.167^{+0.017}_{-0.014}$ & $1.288^{+0.020}_{-0.019}$ \\
~~~~$L_*$\dotfill &Luminosity (\lsun)\dotfill  & $0.2099^{+0.0077}_{-0.0084}$ & $1.047^{+0.055}_{-0.049}$ & $1.789^{+0.066}_{-0.061}$ \\
~~~~$\rho_*$\dotfill &Density (cgs)\dotfill  & $2.898^{+0.085}_{-0.074}$ & $0.842^{+0.029}_{-0.030}$ & $0.756^{+0.060}_{-0.061}$ \\
~~~~$\log{g}$\dotfill &Surface gravity (cgs)\dotfill  & $4.611^{+0.011}_{-0.010}$ & $4.281\pm0.011$ & $4.278^{+0.029}_{-0.033}$ \\
~~~~$T_{\rm eff}$\dotfill &Effective Temperature (K)\dotfill  & $4590\pm49$ & $5404^{+70}_{-67}$ & $5880^{+54}_{-49}$ \\
~~~~$[{\rm Fe/H}]$\dotfill &Metallicity (dex)\dotfill  & $0.18\pm0.08$ & $0.24 \pm 0.09 $ & $0.24 \pm 0.09$ \\
~~~~$Age$\dotfill &Age (Gyr)\dotfill  & $4.1^{+4.3}_{-2.8}$ & $12.82^{+0.73}_{-1.4}$ & $4.7^{+2.7}_{-2.0}$ \\
~~~~$A_V$\dotfill &V-band extinction (mag)\dotfill  & $0.02 \pm 0.01$ & $0.12 \pm 0.08 $ & $ 0.04^{+0.05}_{-0.03}$ \\
~~~~$d$\dotfill &Distance (pc)\dotfill  & $61.89\pm0.08$ & $362.1^{+2.9}_{-2.8}$ & $412.5^{+4.3}_{-4.2}$ \\
~~~~$v \sin i $\dotfill &Projected rotational velocity (\kms)\dotfill  &  $< 2.5 $ &$< 2.5 $ & $< 2.5 $\\
~~~~$P_{\mathrm{rot}}$\dotfill &Rotational period (days)\dotfill  &  $13.2 ^{+1.9} _{-1.4} $ & & \\
~~~~$\log R^{\prime}_\mathrm{HK}$ \dotfill & Ca H\&K chromospheric index (dex) \dotfill& $-4.54 \pm 0.03$ &  $ -4.7 \pm 0.2 $ & $-5.0 \pm 0.3 $ \\

\smallskip\\\multicolumn{2}{l}{Planetary Parameters:}& HIP~65Ab & TOI-157b & TOI-169b\smallskip\\
~~~~$R_P$\dotfill &Radius (\rj)\dotfill  & $2.03^{+0.61}_{-0.49}$ & $1.286^{+0.023}_{-0.020}$ & $1.086^{+0.081}_{-0.048}$ \\
~~~~$M_P$\dotfill &Mass (\mj)\dotfill  & $3.213\pm0.078$ & $1.18^{+0.13}_{-0.12}$ & $0.791^{+0.064}_{-0.060}$ \\
~~~~$P$\dotfill &Period (days)\dotfill  & $0.9809734\pm0.0000031$ & $2.0845435\pm0.0000023$ & $2.2554477\pm0.0000063$ \\
~~~~$T_C$\dotfill &Time of conjunction (\bjdtdb)\dotfill  & $58326.10418\pm0.00011$ & $58326.54771^{+0.00022}_{-0.00021}$ & $58327.44174^{+0.00065}_{-0.00066}$ \\
~~~~$a$\dotfill &Semi-major axis (AU)\dotfill  & $0.01782^{+0.00020}_{-0.00021}$ & $0.03138^{+0.00025}_{-0.00020}$ & $0.03524^{+0.00069}_{-0.00079}$ \\
~~~~$i$\dotfill &Inclination (Degrees)\dotfill  & $77.18^{+0.92}_{-1.00}$ & $82.01^{+0.15}_{-0.16}$ & $80.98^{+0.31}_{-0.38}$ \\
~~~~$b$\dotfill &Transit Impact parameter \dotfill  & $1.169^{+0.095}_{-0.077}$ & $0.8045^{+0.0069}_{-0.0068}$ & $0.9221^{+0.014}_{-0.0098}$ \\
~~~~$e$\dotfill &Orbital eccentricity \dotfill&  0 (adopted, $2\sigma< 0.02$) & 0 (adopted, $2\sigma< 0.21$)& 0 (adopted, $2\sigma< 0.12$) ) \\ 
~~~~$K$\dotfill &RV semi-amplitude (m/s)\dotfill  & $753.7 \pm 5.0 $ & $192\pm20$ & $110.5^{+7.6}_{-6.9}$ \\
~~~~$T_{eq}$\dotfill &Equilibrium temperature (K)\dotfill  & $1411\pm15$ & $1588^{+21}_{-20}$ & $1715^{+22}_{-20}$ \\
~~~~$R_P/R_*$\dotfill &Radius of planet in stellar radii \dotfill  & $0.287^{+0.088}_{-0.068}$ & $0.11329^{+0.00056}_{-0.00054}$ & $0.0866^{+0.0056}_{-0.0031}$ \\
~~~~$a/R_*$\dotfill &Semi-major axis in stellar radii \dotfill  & $5.289^{+0.051}_{-0.045}$ & $5.785^{+0.066}_{-0.069}$ & $5.88^{+0.15}_{-0.16}$ \\
~~~~$\delta$\dotfill &Transit depth (fraction)\dotfill  & $0.082^{+0.058}_{-0.034}$ & $0.01283^{+0.00013}_{-0.00012}$ & $0.00750^{+0.0010}_{-0.00053}$ \\
~~~~$Depth$\dotfill &Flux decrement at mid transit \dotfill  & $0.01094\pm0.00033$ & $0.01283^{+0.00013}_{-0.00012}$ & $0.00733^{+0.00036}_{-0.00037}$ \\
~~~~$\tau$\dotfill &Ingress/egress transit duration (days)\dotfill  & $0.01637^{+0.00013}_{-0.00012}$ & $0.02309^{+0.00083}_{-0.00076}$ & $0.03531^{+0.00077}_{-0.0050}$ \\
~~~~$T_{14}$\dotfill &Total transit duration (days)\dotfill  & $0.03274\pm0.00025$ & $0.08941^{+0.00055}_{-0.00052}$ & $0.0711\pm0.0012$ \\
~~~~$T_{FWHM}$\dotfill &FWHM transit duration (days)\dotfill  & $0.01637^{+0.00013}_{-0.00012}$ & $0.06631^{+0.00065}_{-0.00067}$ & $0.03587^{+0.0048}_{-0.00076}$ \\
~~~~$\rho_P$\dotfill &Density (cgs)\dotfill  & $0.48^{+0.61}_{-0.26}$ & $0.686^{+0.080}_{-0.078}$ & $0.76^{+0.14}_{-0.17}$ \\
~~~~$logg_P$\dotfill &Surface gravity \dotfill  & $3.29^{+0.24}_{-0.23}$ & $3.247^{+0.046}_{-0.050}$ & $3.219^{+0.058}_{-0.079}$ \\
~~~~$\Theta$\dotfill &Safronov Number \dotfill  & $0.072^{+0.022}_{-0.017}$ & $0.0606^{+0.0065}_{-0.0064}$ & $0.0445^{+0.0039}_{-0.0041}$ \\
~~~~$\fave$\dotfill &Incident Flux (\fluxcgs)\dotfill  & $0.899^{+0.038}_{-0.039}$ & $1.446^{+0.077}_{-0.070}$ & $1.964^{+0.10}_{-0.090}$ \\

\hline
\end{tabular}

\end{adjustbox}

\end{minipage}
\end{table*}

\section{Joint analysis of transit light curves and RVs}\label{sec:exofast}
The planetary and stellar parameters for the three systems were modelled jointly and self-consistently using the \tess\ discovery light curves, follow-up photometry and RV measurements from FEROS and CORALIE. We use the most recent version of \exofast\ \citep{Exofastv2, Exofast}, which can fit any number of transits and RV sources while exploring the vast parameter space through a differential evolution Markov Chain coupled with a Metropolis-Hastings Monte Carlo sampler. Built-in Gelman-Rubin statistic \citep{Gelman:1992, Gelman:2003, Ford:2006} is used to check the convergence of the chains. We ran \exofast\ until convergence, and discarded the first chains which have $\chi^2$ above the median $\chi^2$ as the 'burn-in' phase, not to bias the final posterior distributions towards the starting point. 

At each step in the MCMC, we evaluate the stellar properties and limb darkening coefficients by interpolating tables from \cite{ClaretBloemen:2011}. The analytic expressions from \cite{MandelAgol:2002} are used for the transit model and a standard single Keplerian orbit for the RV signal. Four parameters are fitted for the star \teff, \feh, $\log M_*$ and $R_*$. We applied Gaussian priors on \teff\ and \feh\ from the spectral analysis, presented in Sect.~\ref{sec:specMatch}. Stellar density is determined from the transit light curve. The \gaia\ DR2 parallax was used along with SED-fitting of the broad band photometry presented in Table \ref{tab:stellar} to constrain the stellar radius further. We set an upper limit on the V-band extinction from \cite{Schlegel1998} and \cite{Schlafly}, to account for reddening along the line of sight. Combining all this information allows us to perform detailed modelling of the star with the Mesa Isochrones and Stellar Tracks \citep[MIST][]{Dotter2016,Mist1}.

When modelling RVs and transit photometry simultaneously, each planet has five free parameters (assuming a circular orbit) and two additional RV terms for each instrument (CORALIE \& FEROS) for the systemic velocity and RV-jitter. For the transit light curves a set of two limb darkening coefficients for each photometric bands are fitted along with the base line flux and variance of the light curve. The \tess\ PDC-SAP and FFI data were modelled separately to account for different error-properties. For all three planets presented in this study, the precision in the follow-up light-curves is not high enough to detect depth variation as a function of wavelength caused by planetary atmospheric absorption. For the final set of adopted parameters, we fitted one consistent model to all the data which has a fixed transit depth across wavelength.

To avoid Lucy-Sweeney bias of the eccentricity measurement \citep{LucySweeney:1971} we constrain the orbital eccentricity to be zero. To test for possible non-circular orbits, we run a separate MCMC with no constraint on the eccentricity. The data for \Nplanets\ are all consistent with circular orbits. We adopt median values of the posterior distributions and 68\% confidence intervals for the models with eccentricity fixed to zero as the final parameters presented in Table \ref{tab:results}, while quoting the 2 $\sigma$ upper limit of the eccentricity.

HIP 65~A has a star 3.95\arcsec\ away which was not corrected for in the \tess\ light curve. It was also not spatially resolved in the ground-based follow-up photometry. We account for this blending by fitting a dilution parameter for each photometric band, as detailed in Sect. \ref{sec:HIP65B}.

\section{Multi faceted analysis of the HIP~65 system}
HIP~65Ab is an USP Jupiter near the Roche Lobe limit in a binary system and requires an extensive analysis which we present here.

\subsection{Stellar rotation and activity for HIP~65A} \label{sec:HIP65A_rot}
\begin{figure*}
   \centering   
  \includegraphics[width=\textwidth,trim={0cm 0cm 0cm 0cm},clip]{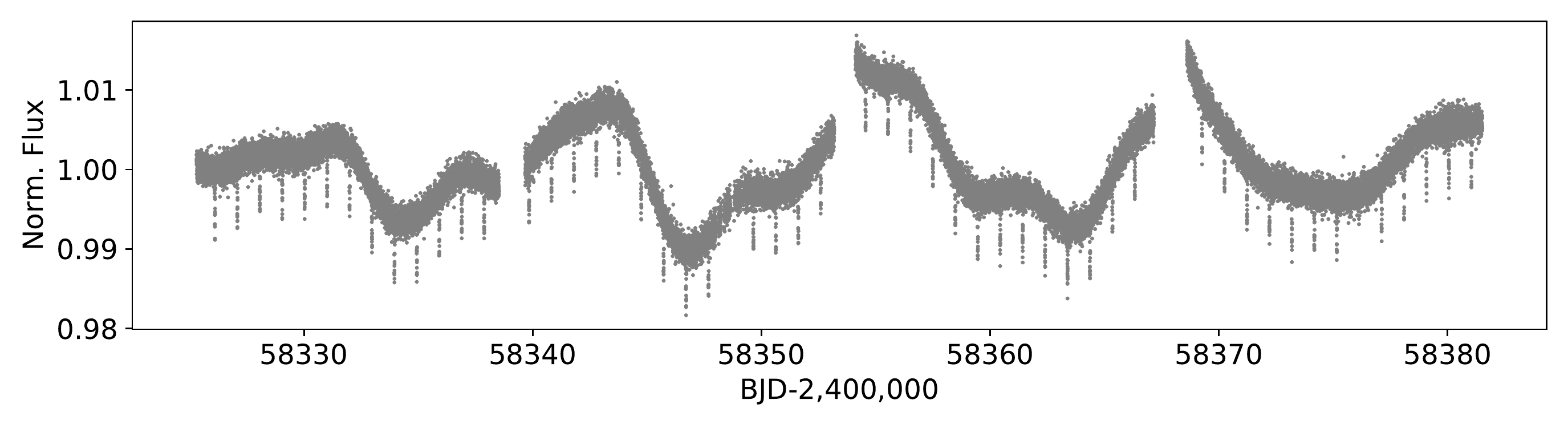}
      \caption{\label{fig:129_LC_TESS} TESS 2-min cadence data for HIP~65A spanning Sectors 1 and 2. The stellar rotational period of $13.2 ^{+1.9} _{-1.4}~\mathrm{days}$ clearly shows up in the PDC-SAP flux. The light curve was flattened while masking the transits before modelling the transits.}
         
\end{figure*}

The PDC-SAP light curve for HIP~65A shows significant stellar variability attributed to star spots coming in and out of view as the star rotates, see Fig. \ref{fig:129_LC_TESS}. Using a Lomb-Scargle periodogram we find a rotation period of $P_{\mathrm{rot}} = 13.2 ^{+1.9} _{-1.4}~\mathrm{days}$. This is in good agreement with the predicted period from $\log R^{\prime}_\mathrm{HK}$ derived in Sect. \ref{sec:specMatch} when using the calibrations from \cite{Suarez:2015}. We find a peak-to-peak variation of about 2~\% which corresponds to a minimum star spot filling factor of $\sim 3$\% of the stellar disc when assuming a sun-like luminosity contrast between spot and continuum as prescribed in \cite{2012A&A...547A..37B} and \cite{2017ApJ...846...99M}.

For the transit analysis presented in Sect. \ref{sec:exofast} we flatten the light curve by fitting third order polynomials to chunks of the light curve while masking the transits. This type of spline filtering acts as a simple low pass filter \citep[see e.g.][]{2016NatAs...1E...4A}. The presence of star spots can affect the radius estimate of transiting planets: 1) as the planet crosses a star spot and the transit shape is thus distorted while the depth is underestimated. 2) the deficit in flux induced by the presence of a cold star spot increases the relative flux blocked by the transiting planet. The later effect leads to an overestimation of the planetary radius. Both of these mechanisms are demonstrated for instance on CoRoT-2 by \cite{2009A&A...504..561W}. For HIP~65Ab these effects are negligible as the uncertainty on the planet radius is dominated by the degeneracy between orbital inclination and planetary radius introduced by the grazing transit configuration. Visual inspection of the transit light curve residuals does not indicate any spot crossing events.

The expected impact of stellar activity for a K-star with $P_{\mathrm{rot}} = 13.2~\mathrm{days}$ on the RVs is of the order of $\sim 10$\ms\ \citep{Suarez:2017,Suarez:2015}. This is comparable to the uncertainty on the FEROS RVs and much smaller than the uncertainties of the CORALIE data. We find no correlation between RV-residuals to the best-fit model and stellar activity indicators, such as bisector span, FWHM of the CCF, $H\alpha$-index. None of the respective Lomb-Scargle periodograms have peaks above 10\% FAP. We do thus not perform any correction for stellar activity.

\subsection{Stellar companion to HIP~65A} \label{sec:HIP65B}

\begin{table}
\centering
\caption{\label{tab:HIP65B} Stellar Properties for HIP~65B, companion to HIP~65A.}
\begin{tabular}{lcc}
	\hline\hline 
	Property	& HIP~65B	 &	Source\\
	\hline
    2MASS ID & None, blended w. HIP~65A             & 2MASS\\
    \gaia\ ID DR2 & 4923860051276772480             & \gaia \\
    TIC  ID &  616112169                          & \tess \\
    \\
    \multicolumn{3}{l}{Astrometric Properties}\\
    R.A.		&	00:00:44.28   	    & \tess	\\
	Dec.			&	-54:49:47.94	    &  \tess	\\
    $\mu_{{\rm R.A.}}$ (\masy)&    $ -207.466 \pm 0.086 $ & \gaia \\
	$\mu_{{\rm Dec.}}$ (\masy)&    $ -72.266 \pm 0.081 $ &  \gaia\\
    Parallax  (mas) &  $16.117 \pm 0.059$  &   \gaia\\
    Distance  (pc)  &  $61.94 \pm 0.23 $    &   \\
    \\
    \multicolumn{3}{l}{Photometric Properties}\\
	V (mag) & $ 16.55 \pm 0.07$ & $\dagger$\\
    G (mag)				&$ 15.3877 \pm 0.0008 $& \gaia \\
    T (mag)	    		&$ 14.30 \pm 0.014 $  &  \tess\\
	\hline
    \end{tabular}
\tablefoot{$\dagger$ V-band magnitude from \cite{2018yCatp047001501K} }
\end{table}

\begin{figure}
\centering
	\includegraphics[width=0.9\columnwidth,trim={11cm 7cm 11cm 7cm},clip]{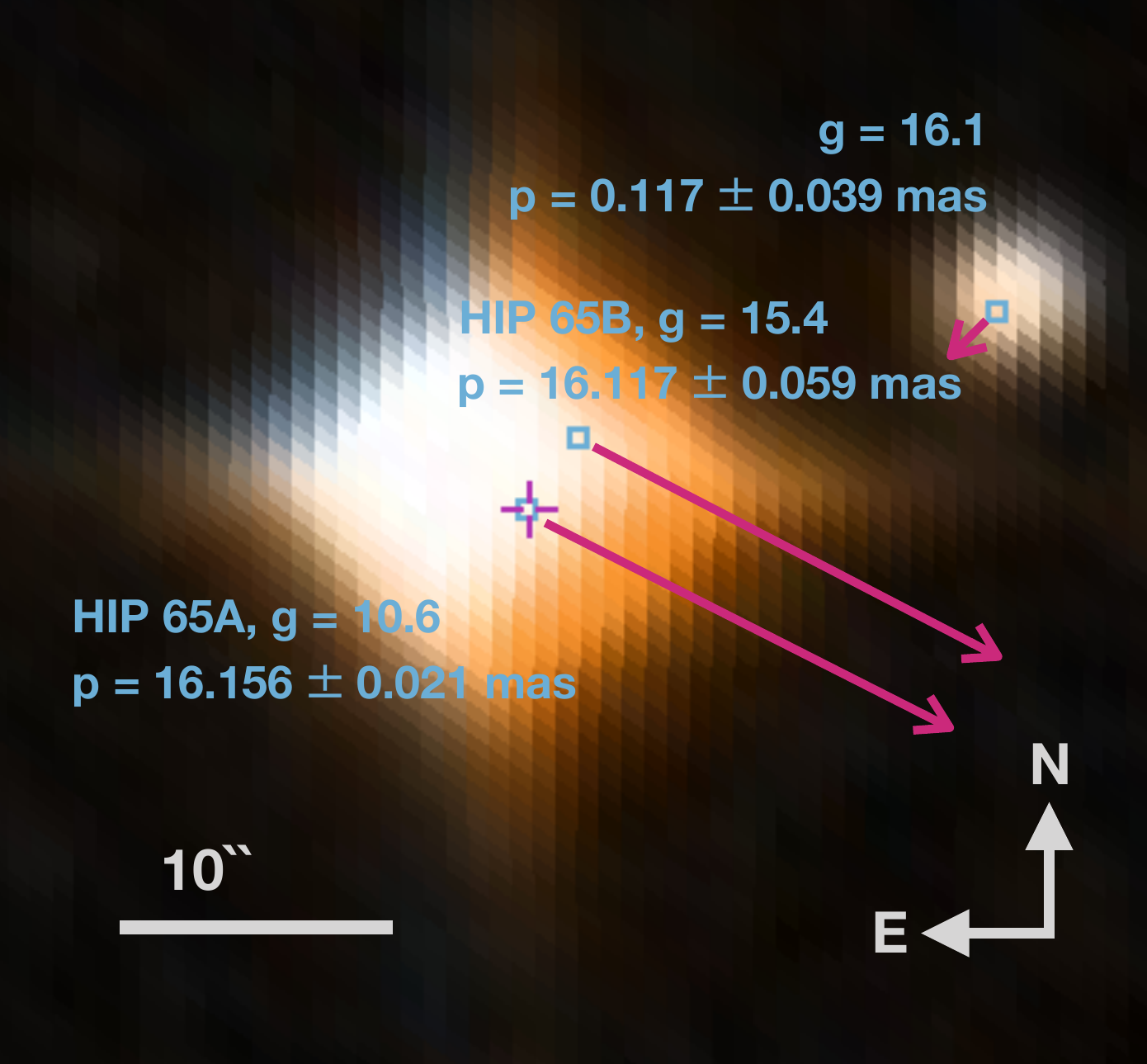}
    \caption{Multi-colour Digitized Sky Survey image of HIP~65A (centre cross-hair) and the nearby companion HIP~65B separated by 3.95\arcsec\ towards north. Their common proper motions are indicated as pink arrows. Blue squares are \gaia\ DR2 sources in the field, with \gaia\ magnitudes and parallaxes denoted.}
    \label{fig:129view}
\end{figure}

HIP~65A is part of a visual binary separated by 3.95\arcsec\ on the sky. The two stars are associated with similar proper motion and parallax \citep{Gaia2018}, as illustrated in Fig. \ref{fig:129view}. We denoted them HIP~65A and HIP~65B. Their angular separation on sky corresponds to 245 AU. HIP~65B is a M-dwarf with \teff\ $= 3713 ^{+994}_{-290} $ K according to \gaia\ DR2. The work by \cite{starHORSE} presents more detailed modelling of \gaia\ stars including HIP~65B. They present a refined effective temperature of $ 3861 ^{+183}_{-259} {\rm K}$ and mass of $0.30 ^{+0.003}_{-0.05} \msun$. Table \ref{tab:HIP65B} summarises the fundamental properties of HIP~65B.

The blending effect from the HIP~65B star was not taken into account when producing the PDC-SAP light curve, as the star was not included in the \tess\ input catalog version 7 \citep[TICv7,][]{TIC7} which was used to correct the normalised light curve for dilution. TICv8 \citep{TIC8} does include HIP~65B which has T = 14.30~mag, which means it is fainter than HIP~65A by $\Delta T = 4.4$~mag. The effect of dilution is small, but non-negligible. Therefore we fitted dilution parameters for this target in all photometric bands, assuming all follow-up light curves include light from both stars. For the \tess\ band we use the \tess\ magnitude to compute the dilution factor. For the photometric bands in which the follow-up light curves were taken we use the Tycho V-band magnitude along with expected magnitude differences from \cite{Pecaut:2013} for a star with the given \teff.

\subsection{Orbital analysis of HIP~65A and HIP~65B using \gaia}

\begin{figure}
\centering
	\includegraphics[width=0.9\columnwidth]{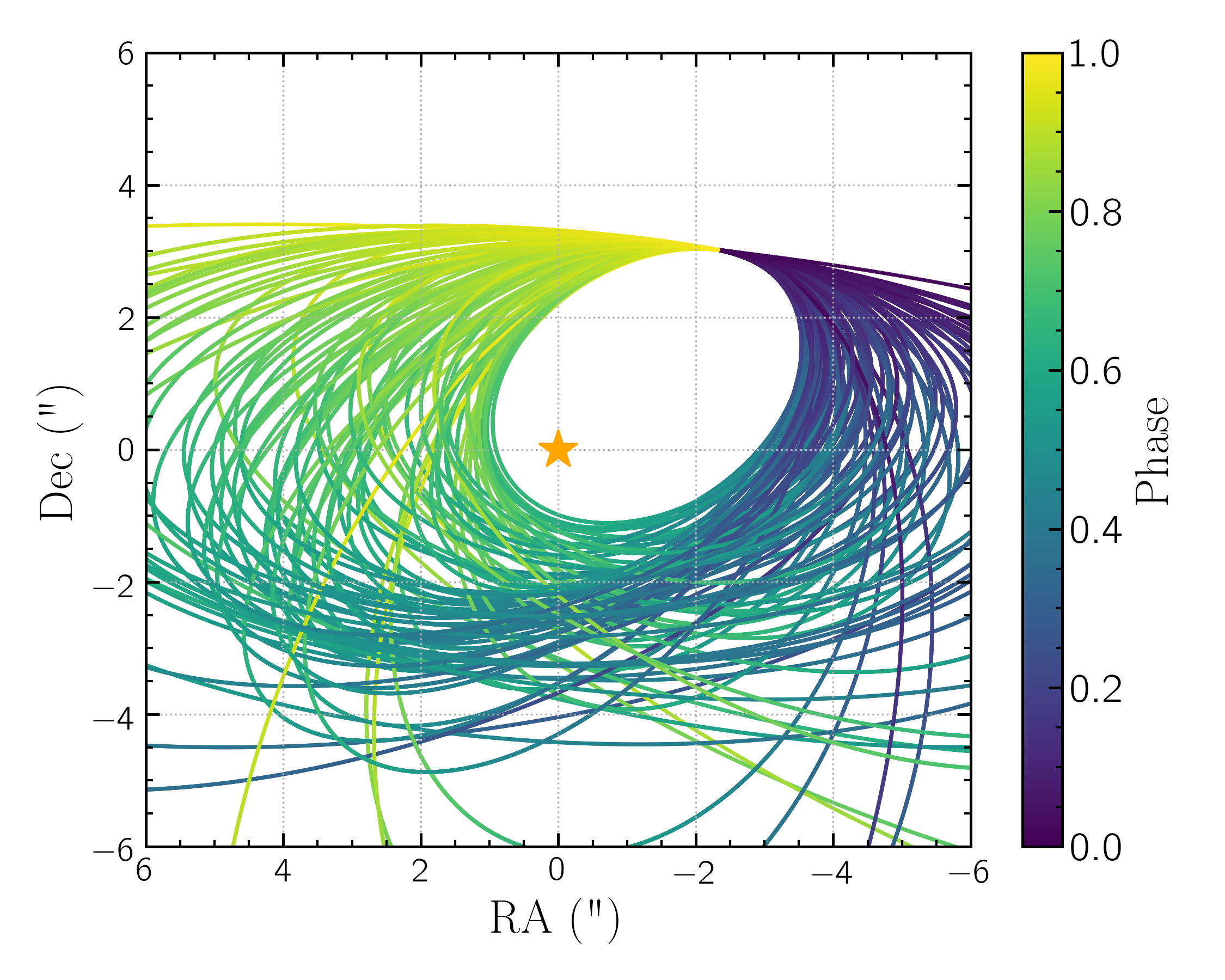}
    \caption{Selection of 100 orbits from the posterior sample of the fit of HIP65B relative to HIP65A using \gaia\ positions and proper motions.  Inclination consistent with HIP65Ab is absent from our posteriors, and low eccentricities are preferred.}
    \label{fig:orbits}
\end{figure}

\gaia\ DR2 measured precise positions and proper motions for HIP~65A and HIP~65B, so we derived orbital element constraints from these measurements using the Linear Orbits for the Impatient algorithm \citep[LOFTI,][]{Pearce2020}. LOFTI uses rejection sampling to determine orbital element posterior probability distributions for stellar binaries derived from \gaia\ DR2 positions and proper motions.  We ran LOFTI on the relative \gaia\ measurements for HIP~65B relative to HIP~65A until the rejection sampling algorithm had accepted 50,000 orbits, comprising our posterior orbit sample.  

The \gaia\ measurements for the pair are not precise enough to constrain the orbital elements to a high degree, as illustrated in Fig. \ref{fig:orbits}. Additionally, HIP~65B has a slightly elevated Renormalised Unit Weight Error (RUWE) of 1.28, whereas RUWE < 1.2 indicates a well-behaved \gaia\ astrometric solution \citep{Lindegren:2018}, so the assumption of a pair of single stars on a Keplerian orbit may not be appropriate. Nevertheless, our results provide some meaningful limits on the orbital architecture of the system, as presented in Table \ref{tab:orbitalparameters}. We find inclinations $109.2^\circ < i < 161.9^\circ$ comprise the majority of the posterior, making edge-on inclination consistent with HIP65Ab highly unlikely.  Low eccentricity ($e < 0.5$) and periastron $>$ 75 AU orbits are preferred.

\begin{table*}
\centering
\caption{\label{tab:orbitalparameters} Orbital Parameter Posterior Distributions for HIP~65A and HIP~65B from \gaia\ Astrometry}
\begin{tabular}{lcccc}
	\hline\hline 
	Parameter$^a$ & Median & Mode & 68\% Min CI$^b$ & 95\% Min CI \\
	\hline
	log$(a)$ (AU) & 2.43 & 2.42 & (2.19, 2.51) & (2.18, 2.82)\\
    $e$ & 0.31 & 0.08 & (0, 0.49) & (0, 0.67) \\
    $i$ ($\circ$)$^c$ & 126.4 & 125.0 & (113.6, 136.5) & (109.2, 161.9)\\
    $\omega$ ($\circ$) & 178.4 & 316.2 & (119.0, 341.8) & (17.9, 360.0)\\
    $\Omega$ ($\circ$)$^d$ & 104.4 & 90.0 & (78.6, 135.1)& (29.6, 178.4)\\
    $T_0$ (yr) & 806.6 & 1319.6 & (-383.7, 1564.9) & (-7192.1, 2013.9)\\
    log$[a\,(1-e)]$ & 2.30 & 2.42 & (1.89, 2.49) & (1.73, 2.71)\\
    \hline
    \end{tabular}
\tablefoot{ (a) Orbital parameters: semi-major axis (a), eccentricity (e), inclination (i), argument of periastron ($\omega$), longitude of nodes ($\Omega$), epoch of periastron passage ($T_0$), and periastron distance [$a\,(1-e)$]\;  (b) Posterior distributions are not Gaussian, so we report the 68\%  and 95\% minimum credible intervals.  (c) Inclination is defined relative to the plane of the sky, $i=90^\circ$ is edge-on.  (d) In the absence of radial velocity information, there is a degeneracy between $\omega$ and $\Omega$, so we limit $\Omega$ to be on the interval [0,180].  If in the future radial velocity is obtained and $\Omega > 180^\circ$, $180^{\circ}$ should be added to both $\Omega$ and $\omega$.}

\end{table*}

\subsection{Phase curve analysis for HIP~65Ab}
\begin{figure} 
   \centering   
  \includegraphics[width=\columnwidth,trim={0cm 0cm 0cm 0cm},clip]{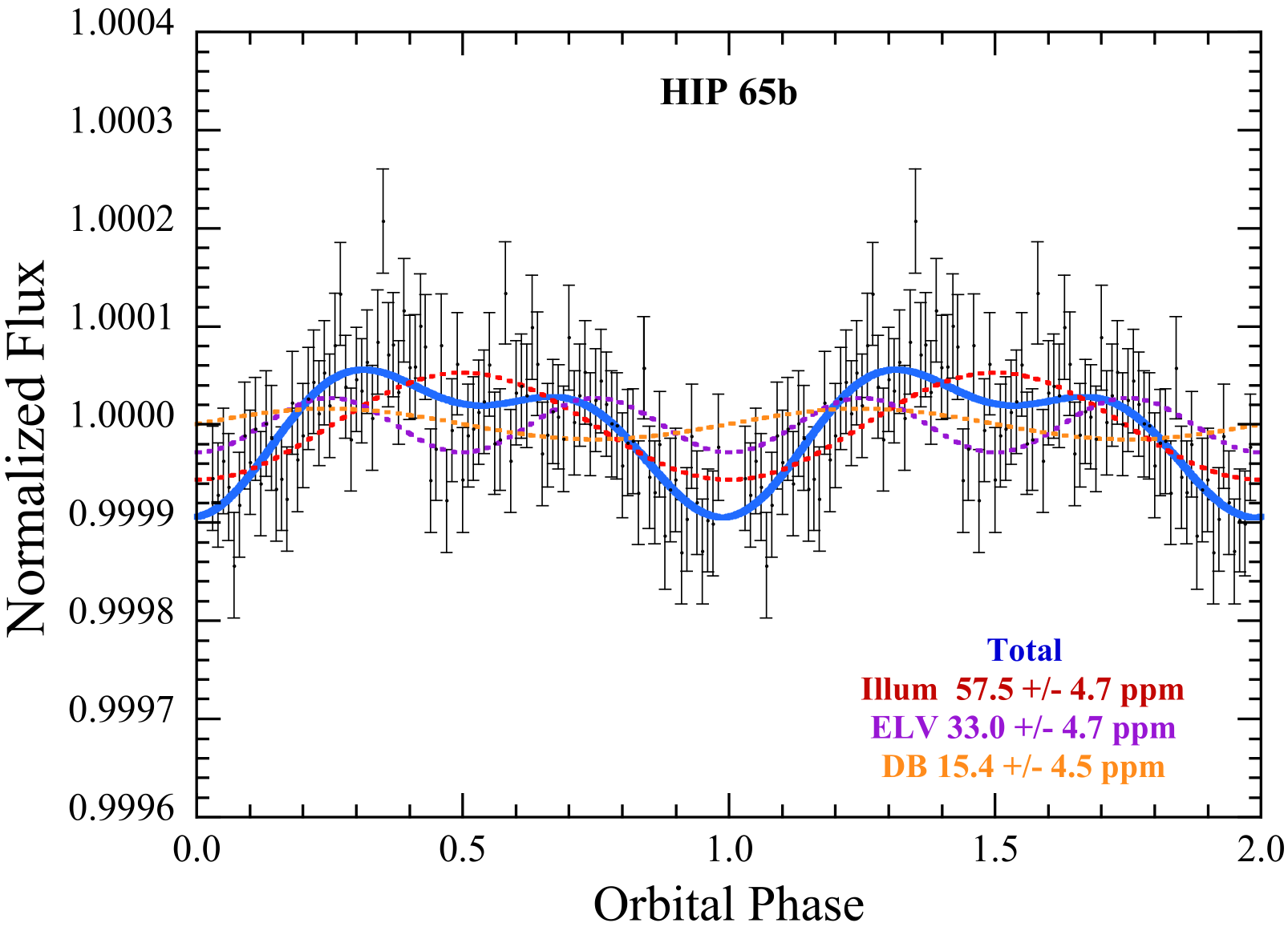}
  \caption{Out-of-transit folded and binned \tess\ light curve for HIP~65Ab.  The red, purple, and orange curves are sinusoids meant to represent the illumination, ellipsoidal light variations, and Doppler boosting effects, respectively (see text for details). The blue curve is the sum of these and represents the best fit of this model to the out-of-transit light curve.}
         \label{fig:phase}
\end{figure}

A \tess\ phase folded light curve of HIP~65Ab is shown in Fig.~\ref{fig:phase} with the eclipses removed.  The data are phase-folded with the orbital period and averaged into 100 bins that are $\sim$14 minutes long, each with the contributions of about 350 individual flux measurements.  For the individual flux measurements, we measure an rms scatter in the data points of $\simeq$ 980 ppm, and thus the statistical uncertainty in each bin of the light curve is approximately 53 ppm.  A casual inspection shows that the light curve exhibits a characteristic orbital phase curve as it has been possible to detect for exoplanets since the \corot\ space-mission \citep{snellen2009,mazeh2010} .  

We fitted sines and cosines of $\omega t$ and $2 \omega t$ to the out-of-transit light curve, where $\omega$ is the angular frequency of the orbit, to represent various physical effects \citep[see e.g.][]{2010ApJ...715...51V,2011ApJ...728..139C,2017PASP..129g2001S,2018arXiv181209227N,2019AJ....157..178S}.  We limited ourselves to just these four terms given the limited statistics in our folded out-of-eclipse light curve.  The red curve in Fig.~\ref{fig:phase} is the $\cos \omega t$ term representing the illumination effect of the host star on the planet; the purple curve is the $\cos 2\omega t$ term to approximate the bulk of the ellipsoidal variations (`ELVs'); and the orange curve is the $\sin \omega t$ term for the Doppler boosting effect \citep{2003ApJ...588L.117L,2010ApJ...715...51V}. The three terms were detected at the 12, 7, and 3.4 $\sigma$ confidence levels, respectively, and there was no statistically significant amplitude for a $\sin 2\omega t$ term, where no physical effect is expected.

We next utilised the amplitudes of the ELV and Doppler boosting terms to make an independent determination of the planetary mass. Following the expressions and references in \cite{2019AJ....157..178S} we adopted a Doppler boosting coefficient in front of the $K_{RV}/c \sin i$ term of $4.2^{+1.8}_{-1.2}$ and an ELV coefficient in front of the $q (R_p/a)^3 \sin^2i$ term of $1.25 \pm 0.25$, where $K_{RV}$ is the orbital RV semi-amplitude of the host star, $q$ the planet to host star mass ratio, and $a$ is the orbital radius of the planet.  Since we know the mass of the host star and the orbital inclination to $\sim  1^{\circ}$, either the Doppler boosting or ELV measurement, in principle, determines the planetary mass. We therefore carried out a Monte Carlo evaluation of the overall uncertainty in the planet mass using both measurements \citep{1984ARA&A..22..537J}. From this analysis we find $M_p = 3.4 \pm 0.6\ \mj$, which is in agreement with RV-derived mass of $3.213\pm0.078\ \mj$.

Finally, in regard to the out-of-transit light curve of HIP~65Ab, we explored what we can learn from the illumination term which has an amplitude of 57 ppm.  Because the estimated equilibrium temperature of the planet at the sub-stellar point is likely $\lesssim 1400$ K, we neglect any contribution from the thermal emission of absorbed and reprocessed radiation from the host star.  We find that if the Bond albedo of the facing hemisphere of the planet is allowed to be in the range of $0-0.5$, then the resultant likelihood distribution of planet radii, as inferred from the illumination term, is close to 1 \rj. On the other hand, if the geometric albedo is constrained to be $\lesssim 0.1$, then the peak of the radius distribution is close to our transit-based estimate of 2 \rj. This low albedo is quite consistent with the results found recently for WASP-18b \citep{2019AJ....157..178S}.


\section{Results and discussion}
For each system we list the final stellar and planetary parameters in Table \ref{tab:results} with 1 $\sigma$ errors. Figures \ref{fig:129_FUP} through \ref{fig:169_RV} show the final joint model fitted to the discovery and follow-up data.

\subsection{HIP~65Ab}
HIP~65Ab is an ultra short period ($P=0.98$~days) Jupiter with mass $3.213 \pm 0.078\ \mj$. Its radius, $R=2.03 ^{+0.61}_{-0.49}\ \rj$, is poorly constrained as the transit is extremely grazing with impact parameter $b=1.169^{+0.095}_{-0.077}$. The planet is thus barely transiting with less than half its disc covering the host star during transit.  Determining the stellar limb darkening is especially important for a grazing transit where the planet never leaves the limb. In the case of HIP~65A we derive linear and quadratic limb-darkening coefficients  $u_1 = 0.545\pm0.037$ and $u_2 = 0.195\pm0.041$ for the \tess\ band.
The main source of the uncertainty on the planetary radius is the degeneracy between the orbital inclination of the planet and its radius.

\begin{figure} 
   \centering   
  \includegraphics[width=\columnwidth,trim={0cm 0cm 0cm 0cm},clip]{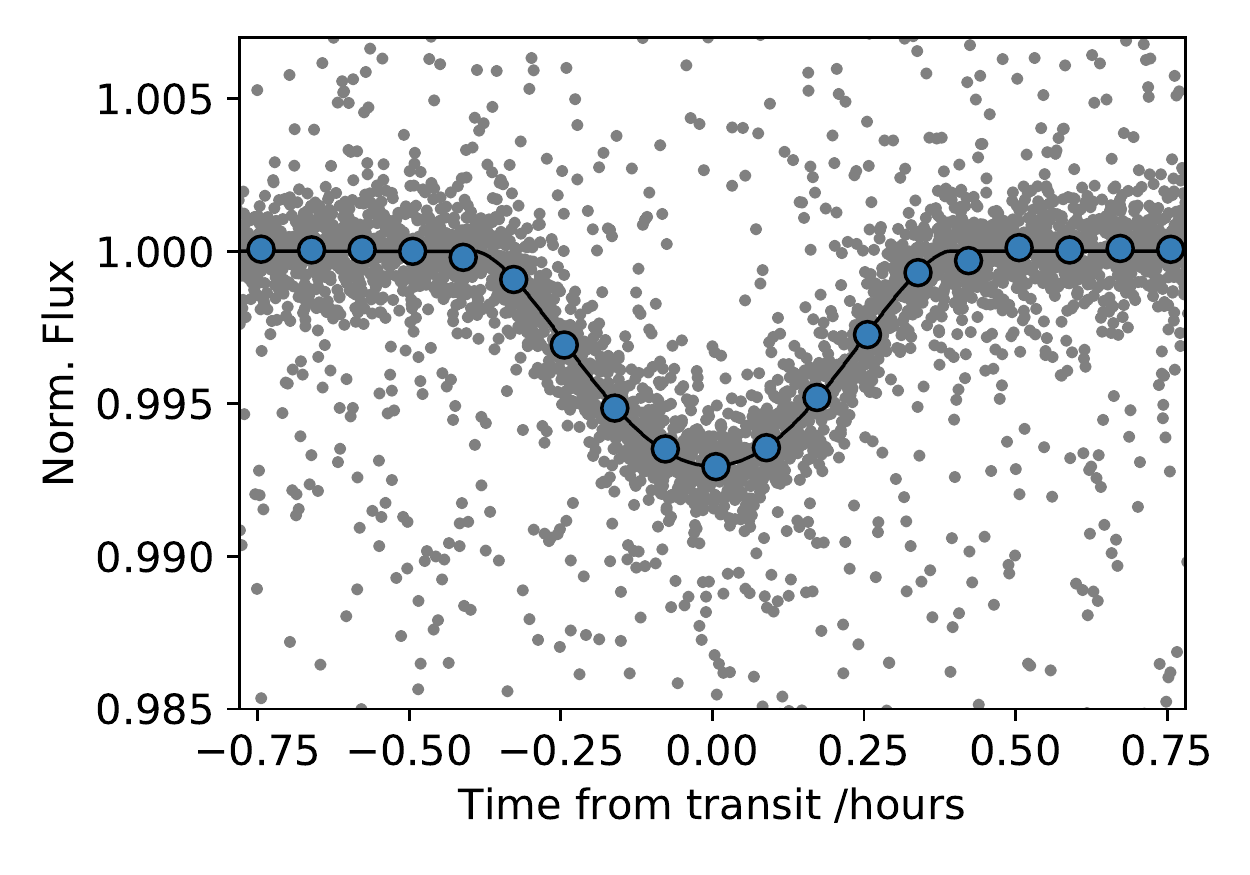}
      \caption{\label{fig:129_LC} Phase folded transit light curve for HIP~65Ab including data from two \tess\ Sectors (see Fig. \ref{fig:129_LC_TESS} for the full \tess\ light curve) and ground based follow-up photometry. Individual follow-up light curves are shown in Fig. \ref{fig:129_FUP}. Grey points are the un-binned data. The points with large scatter come from the NGTS 10 sec cadence observations. In blue are the data binned to 5 min in phase space. }
\end{figure}

The V-shaped, relatively shallow, transit model can be seen in Figs. \ref{fig:129_FUP} and \ref{fig:129_LC} plotted along with the follow-up light curves and \tess\ data. Figure \ref{fig:169_RV} shows the phase folded RVs showing the large semi-amplitude of $754 \pm 5$~ \ms.

HIP~65A is a bright (V = 11.1) main sequence K-star with \teff\ $ = 4590\pm 49~ \mathrm{K}$, $R_* = 0.724 \pm 0.009\ \rsun$ and $M_* = 0.781 \pm 0.027\ \msun$. We find clear signs of stellar rotation in the \tess\ light curve corresponding to a rotation period of $P_{\mathrm{rot}} = 13.2 ^{+1.9} _{-1.4}~\mathrm{days}$. The peak-to-peak modulation of the light curve is consistent with a 3\% minimum filling factor of star spots on the stellar surface of HIP~65A. This is much higher than for our own Sun, but consistent with other active K-dwarfs, such as the canonical planet host HD~189733 \citep{2011MNRAS.416.1443S}. For moderately rotating main sequence K- and G-stars we expect spots to be located towards the equator of the star \citep{1996A&A...314..503S}. The lack of star spot crossings seen in our data could be indicating that HIP~65Ab transits one of the stellar poles in a co-planar orbit.

HIP~65A has an associated stellar companion, HIP~65B, separated by 3.95\arcsec\ with similar distance and proper motion. Based on \gaia\ DR2 data we conclude that HIP~65B is an M-dwarf separated by 269 AU.  
With such a separation and high mass ratio $q=0.38$ the protoplanetary disc is not expected to be affected by the presence of the stellar companion \citep{Artymowicz1994,Patience2008}.

The orbital analysis of HIP~65A+B using \gaia\ measurements indicates that the mutual inclination is less than 0.5. This still includes orbital solutions where the Lidov-Kozai mechanism is invoked, which could be used to explain the architecture of the system \citep{Lidov1962,Kozai1962}. A requirement for such a process to occur is that the mutual inclination between the two orbits at high period ratio is large enough. From then, the angular momentum exchange between the two orbits will induce phase-opposed oscillations of the eccentricity and inclination of the inner orbit. At high eccentricity phases, tidal dissipation will take place during the periastron passages, leading the orbit of the planet to shrink. This mechanism was already successfully introduced to explain the observations of planets in binary systems \citep[e.g.][]{Wu2003,Fabrycky2007}.

Measuring the spin-orbit misalignment between the central star and the inner planet could help at selecting the mechanism responsible for the current architecture of the system. If a significant misalignment is found, the Lidov-Kozai mechanism will be favoured. If, on the other hand, the spin axis of the star is aligned with the normal to the inner orbit, then the Lidov-Kozai mechanism will be excluded because the planet would not have been misaligned from its original orbit. 

\begin{figure} 
   \centering   
  \includegraphics[width=\columnwidth,trim={0cm 0cm 1cm 0cm},clip]{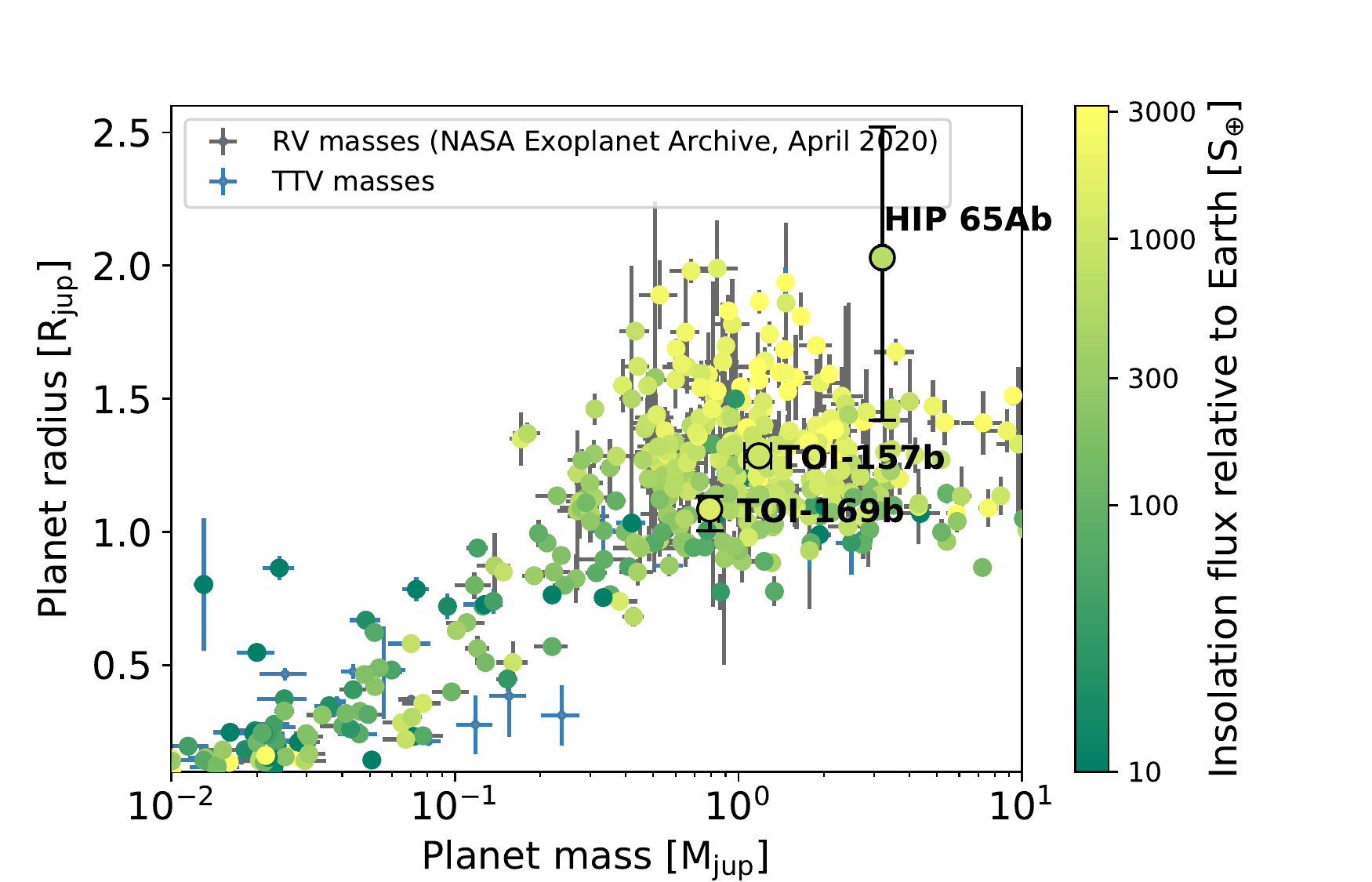}
      \caption{\label{fig:MR} Mass and radius for known exoplanets extracted from NASA Exoplanet Archive. Only planets with 20\% precision on their mass are included. \Nplanets\ are plotted in blue. }
\end{figure}

\begin{figure} 
   \centering   
  \includegraphics[width=\columnwidth,trim={0cm 0cm 1cm 0cm},clip]{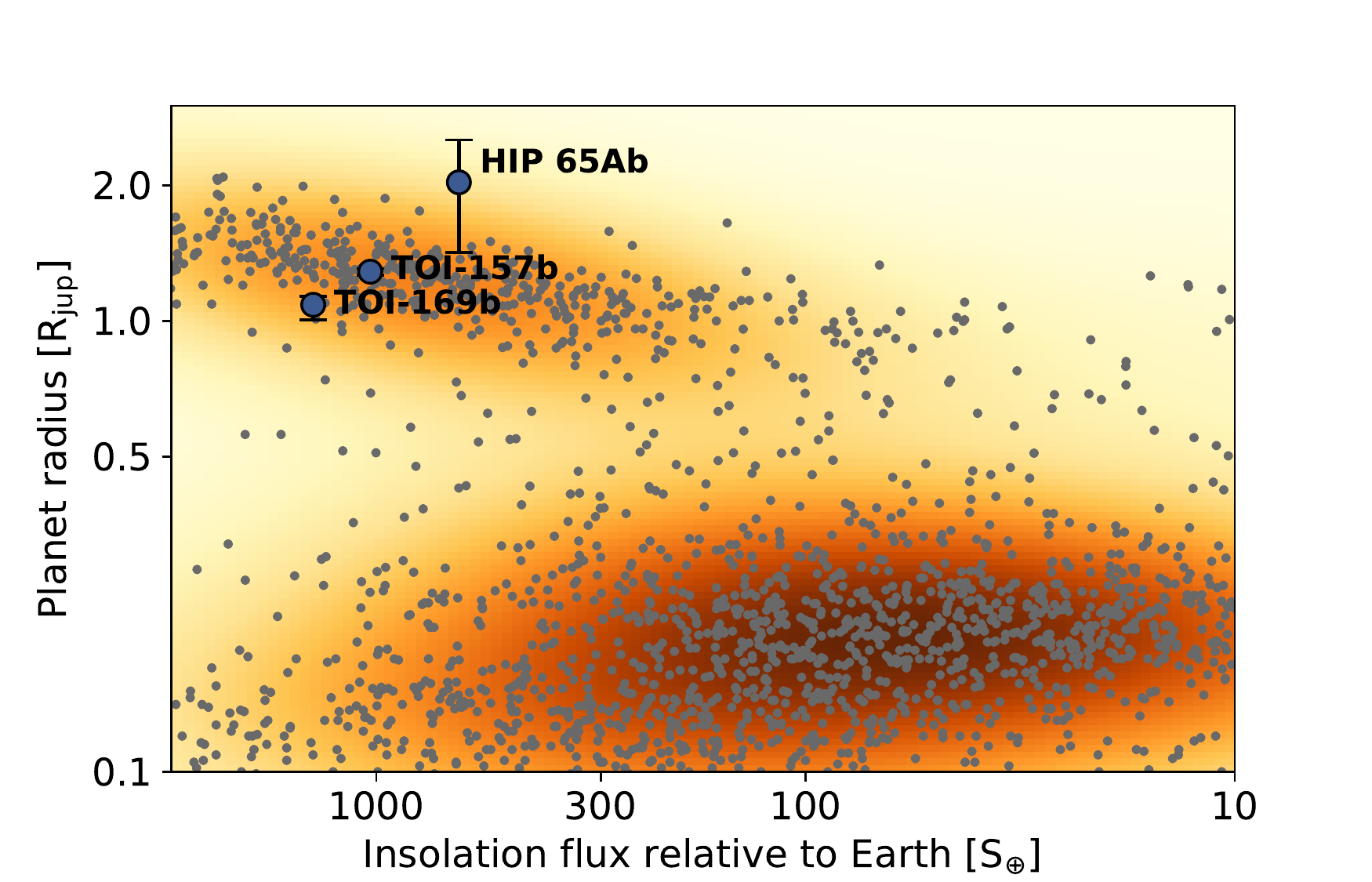}
      \caption{\label{fig:insolation} Insolation flux relative to Earth plotted against radii for known exoplanets extracted from NASA Exoplanet Archive. The orange contours indicate point density (not occurrence) \Nplanets\ are plotted in blue. }
\end{figure}

Figure~\ref{fig:MR} shows mass and radius for known exoplanets with \Nplanets\ over-plotted in blue. HIP~65Ab does appear to have an unusually large radius which most likely is overestimated due to the grazing nature of the transit. Close-in gas planets are found to be inflated, as the proximity to the host star can inhibit thermal contraction \citep{2010RPPh...73a6901B, 2010ApJ...714L.238B}. As seen in Fig. ~\ref{fig:insolation} HIP~65Ab receives 642 times more insolation flux than that of the Earth. Given the mass and insolation flux it is unlikely that HIP~65Ab is larger than 1.5 \rj.

The effects of the large planetary mass and radius, relative to the host star, are evident in the \tess\ light curve. Our analysis of the phase curve yields an illumination effect amplitude of $57.5 \pm 4.7~\mathrm{ppm}$, ELV amplitude $30.0 \pm 4.7~\mathrm{ppm}$, as well as Doppler boosting effect $15.4 \pm 4.5~\mathrm{ppm}$. The mass derived on the basis of the two latter terms is $3.4 \pm 0.6\ \mj$, in agreement with the independently derived RV-mass. We estimate the geometric Bond albedo to be $\lesssim 0.1$, but cannot constrain it further due to large uncertainties on the radius. A study by \cite{Wong:2020} presents a systematic phase curve analysis of TOIs for the first year of \tess\ operation, which are in agreement with our results.

The tidal interaction between HIP~65Ab and its host star is expected to spin up the stellar rotation while removing angular momentum from the orbit. Over time the orbit will circularise and the planet will spiral within the Roche limit of HIP~65A and disintegrate. 
We compute the Roche limit, $a_{Roche}$, as defined for a  infinitely compressible object in \cite{faber2004}:
\begin{equation}
a_{Roche} = 2.16 R_p \left( \frac{M_s}{M_p} \right) ^{1/3},
\end{equation}
where $R_p$ and $M_p$ are the planet radius and mass respectively and $M_s$ the stellar mass. The Roche limit for HIP~65Ab is 0.013 AU when using using the values listed in Table \ref{tab:results}. If using the more realistic planet radius of 1.5 \rj\ the resulting Roche limit is 0.010 AU. This means that HIP~65Ab is orbiting its host star at a distance corresponding to less than twice the Roche limit. 

The efficiency of the tidal dampening is given by the \emph{stellar reduced tidal quality factor} $Q^{\prime}_s \equiv 3/2~ Q_s / k_2$, where $ Q_s $ is the tidal quality factor and $k_2$ the second-order potential Love number. $Q^{\prime}_s$ can vary from $10^5$ to $10^9$ and depends on stellar properties which will change throughout the lifetime of the system \citep{2007ApJ...661.1180O,2016A&A...589A..55D,2018AJ....155..165P}. 

We calculated the remaining lifetime $t_{remain}$ of the planet using the prescription for slowly rotating stars in \cite{2011MNRAS.415..605B}:
\begin{equation}
t_{remain} = \frac{2 Q^{\prime}_s }{17 n} \frac{M_s}{M_p} \left( \frac{a}{R_s} \right) ^5,~~ n=\sqrt{\frac{G (M_s + M_p)}{a^3}},
\end{equation}
where $a$ is the semi-major axis of the planetary orbit and $R_s$ the stellar radius. For HIP~65Ab $t_{remain}$ is 76 Myr when using $Q^{\prime}_s = 10^7$ and $7.6 ~\mathrm{Gyr}$ for $Q^{\prime}_s = 10^9$. This is much shorter than the expected age derived from the global modelling using MIST, $4.1^{+4.3}_{-2.8}$ Gyr. 

Gyrochronology yields an age of $0.32^{+0.1}_{-0.06} \mathrm{Gyr}$ \citep{Barnes2007}. The discrepancy between the two age estimates indicates that the star has been spun up, making the approach of gyrochronology unfeasible. In order to get a life span of the system that is consistent with the MIST age one must use  $Q^{\prime}_s > 10^8$. Other systems with short period planets, such as HAT-P-11b \citep{2010ApJ...710.1724B}, show evidence of tidal spin which induces increased stellar activity \citep{2017ApJ...848...58M}.

\begin{figure} 
   \centering   
  \includegraphics[width=\columnwidth,trim={0cm 0cm 0cm 0cm},clip]{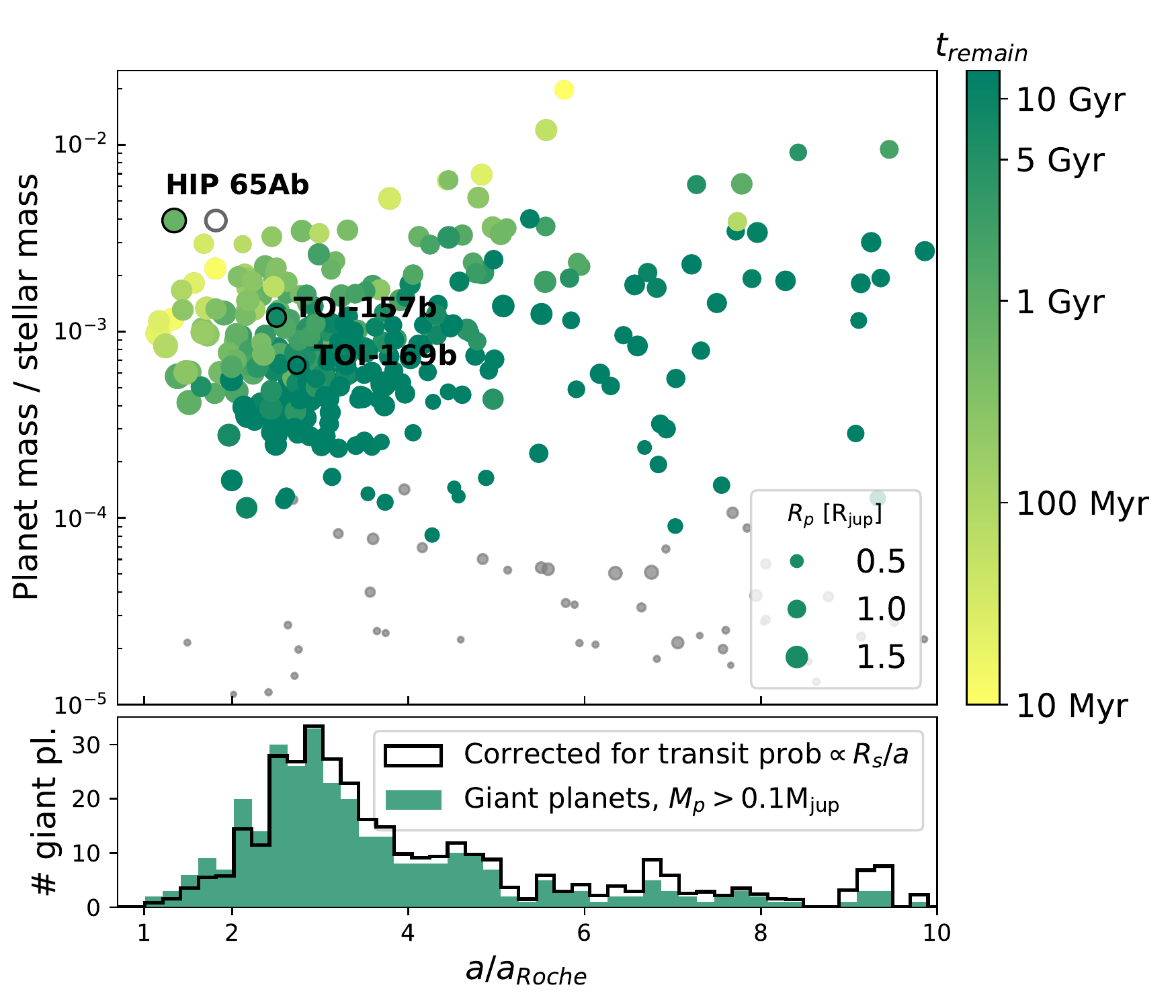}
      \caption{\label{fig:roche} Top panel: Orbital separation in units of Roche limit as a function of the planet-to-star mass ratio for the planets as in Fig. \ref{fig:MR}. The symbols sizes are proportional to the planer radius and the colour-coding represents $t_{remain}$ when assuming  $Q^{\prime}_s = 10^7$. The open grey circle is HIP~65Ab when assuming a radius of 1.5 \rj. Filled grey circles are planets with mass < 0.1 \mj, which are excluded from the histogram in the bottom panel. 
      Bottom panel: Distribution of $a/a_{Roche}$ for a subset of the planets from the top and Fig. \ref{fig:MR} with masses above 0.1 \mj. The green histogram shows the raw count of planets in bins of 0.2. The solid line is the same distribution but weighted by the inverse of the transit probability for each planet.}
      
\end{figure}

Figure \ref{fig:roche} shows the orbital separation of known exoplanets normalised with the Roche limit as a function of planet-to-star mass ratio. The symbol sizes are proportional to the planet radius and the colour-coding represents $t_{remain}$ when assuming  $Q^{\prime}_s = 10^7$. HIP~65Ab is bordering the empty parameter space representing massive planets close to the Roche limit. The dearth of targets could be the consequence of Jupiters spiralling into their host star \citep{2018MNRAS.476.2542C}. The bottom panel shows a histogram of the orbital separation in units of $a_{Roche}$, for giant planets only (\mpl > 0.1 \mj).  It is evident that several known exoplanet have $a/a_{Roche} < 2$ and subsequent short predicted remaining lifetimes. The distribution which peaks at $a/a_{Roche} \sim 3 $ has been analysed before by \cite{2017A&A...602A.107B} amongst others. This could be an artefact caused by planets migrating inwards from highly eccentric orbits through tidal dissipation \citep{2006ApJ...638L..45F}. Such planets would subsequently circularise which would lead to a bunch up at ~$3 a_{Roche}$. Disc driven migration on the other hand would result in an inner build up precisely at the Roche limit \citep{1998Sci...279...69M}.


\subsection{TOI-157b}
TOI-157b is an inflated hot Jupiter with orbital period $P=2.08$~days, mass $1.18 \pm 0.13\ \mj$ and $R=1.29 \pm 0.02\ \rj$. The photometry from the ground-based follow-up and \tess\ is presented in Fig. \ref{fig:157_LC}. The RVs are shown in Fig. \ref{fig:157_RV}, including two FEROS RVs which where not used in the analysis in the end, as they do not help constrain the amplitude of the RV curve when fitting an offset between CORALIE and FEROS. 

The host star TOI-157 is a slightly evolved G-type sub-giant with \teff\ $= 5398 \pm 67~ \mathrm{K}$, $R_* = 1.17 \pm 0.02\ \rsun$ and $M_* = 0.95 \pm 0.02\ \msun$. Through modelling the star with MIST we compute an age of $12.9^{+0.69}_{-1.4} \mathrm{Gyr}$. Given the evolved nature of TOI-157, the planet receives a considerable amount of insolation flux; 1032 times that of Earth, corresponding to an equilibrium temperature of $1588 \pm 20 \mathrm{K}$. TOI-157b has a separation of just 0.03 AU to its sub-giant host star. Planets orbiting close-in ($a<0.5$ AU) to evolved stars are very rare \citep{frink01,johnson10,jones11} though \tess\ has provided several new detections around subgiants \citep[TOI-120b, TOI-172b and TOI-197b:][]{nielsen2019,brahm2019,2019AJ....157..191R,2019AJ....157..245H,2019AJ....157...51W}.


\subsection{TOI-169b}
TOI-169b has the longest period of the three planets presented in this study with $P=2.26$~days. It is a low-mass hot Jupiter with mass $0.79 \pm 0.06\ \mj$ and radius $R=1.086 ^{+0.081}_{-0.048}\ \rj$. TOI-169 is found to be a main sequence G1-star with \teff\ $= 5880 \pm 50~ \mathrm{K}$, $R_* = 1.288 \pm 0.020\ \rsun$ and $M_* = 1.1477 ^{+0.069}_{-0.075}\ \msun$. 

Despite having the longest orbital period of the three planets presented in this study, TOI-169b receives the highest insolation flux; 1403 times that of Earth, corresponding to an equilibrium temperature of $1715 \pm 21 \mathrm{K}$. Figure \ref{fig:insolation} shows the known population of exoplanets plotted in insolation-radius space. TOI-169b is located right at the edge of the Neptune desert. Given its irradiation, TOI-169 is unusually dense, which could support a scenario of the atmospheric volatile layer being stripped away by photo-evaporation, to a point where the self-gravity of the planet is strong enough to withstand the atmospheric escape \citep{2014ApJ...792....1L, 2015IJAsB..14..201M}. During this process, less massive planets could completely lose their outer layer and end up as a naked core at the bottom of the desert \citep{2018MNRAS.479.5012O}, thus joining the large population of mainly \kepler\ planets seen in Fig. \ref{fig:insolation}.


\section{Conclusions}
We have presented the discovery and mass determination of three new Jovian planets \Nplanets\ from the \tess\ mission. We based our analysis on both 2-min cadence and FFI data from \tess\ spanning multiple Sectors in the first year of operations as well as numerous ground-based photometric observations. Light curves were modelled jointly with RVs from the CORALIE and FEROS spectrographs. Using SOAR speckle imaging we rule out close stellar companions for all three host stars.

HIP~65Ab is an ultra short period massive hot Jupiter with a period of 0.98 days, orbiting one component of a stellar binary. Despite the proximity to its host star, HIP~65Ab receives the least amount of radiation out of the three planets presented in this study. We find evidence that HIP~65Ab is spinning up its host star though tidal interaction. The planet's semi-major axis is less than twice the separation at which it would be destroyed by Roche lobe overflow. The predicted remaining lifetime ranges from 80 Myr to a few Gyr, assuming a reduced tidal dissipation quality factor of $Q^{\prime}_s = 10^7 - 10^9$. TOI-157b and TOI-169b both receive more than 1000 times the Earth's insolation flux. TOI-157b orbits a sub-giant star with a 0.03 AU separation. TOI-169b is bordering the Neptune desert and can thus help solve the conundrum of which mechanisms are responsible for the shortage of close-in giant planets.


\begin{acknowledgements}
We  thank  the  Swiss  National  Science  Foundation  (SNSF) and the Geneva University for their continuous support to our planet search programmes. This work has been in particular carried out in the frame of the National Centre for Competence in Research ‘PlanetS’ supported by the Swiss National Science Foundation (SNSF). 
This publication makes use of The Data \& Analysis Center for Exoplanets (DACE), which is a facility based at the University of Geneva (CH) dedicated to extrasolar planets data visualisation, exchange and analysis. DACE is a platform of the Swiss National Centre of Competence in Research (NCCR) PlanetS, federating the Swiss expertise in Exoplanet research. The DACE platform is available at \url{https://dace.unige.ch}. 
This paper includes data collected by the \tess\ mission. Funding for the \tess\ mission is provided by the NASA Explorer Program. 
Resources supporting this work were provided by the NASA High-End Computing (HEC) Program through the NASA Advanced Supercomputing (NAS) Division at Ames Research Center for the production of the SPOC data products.\\ 
This work has made use of data from the European Space Agency (ESA) mission
\gaia\ (\url{https://www.cosmos.esa.int/gaia}), processed by the \gaia\
Data Processing and Analysis Consortium (DPAC,
\url{https://www.cosmos.esa.int/web/gaia/dpac/consortium}). Funding for the DPAC has been provided by national institutions, in particular the institutions participating in the \gaia\ Multilateral Agreement.
This research has made use of \emph{Aladin sky atlas} developed at CDS, Strasbourg Observatory, France. This work makes use of observations from the LCOGT network.
This work is partly supported by JSPS KAKENHI Grant Numbers JP15H02063, JP18H01265, JP18H05439, JP18H05442, and JST PRESTO Grant Number JPMJPR1775.
The IRSF project is a collaboration between Nagoya University and the South African Astronomical Observatory (SAAO) supported by the Grants-in-Aid for Scientific Research on Priority Areas (A) (Nos. 10147207 and 10147214) and Optical \& Near-Infrared Astronomy Inter-University Cooperation Program, from the Ministry of Education, Culture, Sports, Science and Technology (MEXT) of Japan and the National Research Foundation (NRF) of South Africa.
We thank Akihiko Fukui, Nobuhiko Kusakabe, Kumiko Morihana, Tetsuya Nagata, Takahiro Nagayama, Taku Nishiumi, and the staff of SAAO for their kind supports for IRSF SIRIUS observations and analyses.
The research leading to these results has received funding from  the ARC grant for Concerted Research Actions, financed by the Wallonia-Brussels Federation. TRAPPIST is funded by the Belgian Fund for Scientific Research (Fond National de la Recherche Scientifique, FNRS) under the grant FRFC 2.5.594.09.F, with the participation of the Swiss National Science Fundation (SNF). MG and EJ are F.R.S.-FNRS Senior Research Associate.
R.B.\ acknowledges support from FONDECYT Post-doctoral Fellowship Project 3180246, and from the Millennium Institute of Astrophysics (MAS).
A.J.\ acknowledges support from FONDECYT project 1171208 and by the Ministry for the Economy, Development, and Tourism's Programa Iniciativa Cient\'{i}fica Milenio through grant IC\,120009, awarded to the Millennium Institute of Astrophysics (MAS). 
JSJ acknowledge support by FONDECYT grant 1161218 and partial support from CONICYT project Basal AFB-170002.
JVS and LAdS are supported by funding from the European Research Council (ERC) under the European Union's Horizon 2020 research and innovation programme (project {\sc Four Aces}; grant agreement No 724427).  
L.A.P. is supported by the National Science Foundation Graduate Research Fellowship Program under Grant No. DGE-1746060. Any opinions, findings, and conclusions or recommendations expressed in this material are those of the author(s) and do not necessarily reflect the views of the National Science Foundation.
Includes data collected under the NGTS project at the ESO La Silla Paranal Observatory.  The NGTS facility is funded by
the University of Warwick,
the University of Leicester,
Queen's University Belfast,
the University of Geneva,
the Deutsches Zentrum f\" ur Luft- und Raumfahrt e.V. (DLR; under the `Gro\ss investition GI-NGTS'),
the University of Cambridge
and the UK Science and Technology Facilities Council (STFC; project references ST/M001962/1 and ST/S002642/1).
Staff from the University of Warwick acknowledge support from STFC consolidated grant ST/P000495/1.
\end{acknowledgements}

\bibliographystyle{aa} 
\bibliography{USP}

\begin{appendix} 

\section{RV data} \label{sec:RVdata}

\begin{table}[ht]
\caption{\label{tab:129rvs}Radial velocity measurements from CORALIE and FEROS for HIP~65A.}
\centering                          
\begin{tabular}{l c c c c}        
\hline\hline                 
BJD & RV  & $\sigma_{\mathrm{RV}}$ & BIS  & Instrument \\
(- 2,400,000) & (\ms) & (\ms) &  (\ms) &  \\
\hline                        
~58382.793749	& 22105.0 &	25.5 &	-40.7	 & CORALIE \\
~58406.764434	& 20526.8 &	66.6 &	-118.2	 & CORALIE \\
~58408.788981	& 20652.1 &	28.6 &	-134.2	 & CORALIE \\
~58410.498946	& 21177.3 &	31.6 &	-64.6	 & CORALIE \\
~58410.586650	& 20825.2 &	31.5 &	-49.1	 & CORALIE \\
~58410.732244	& 20533.6 &	31.2 &	21.7	& CORALIE \\
~58411.498626	& 21082.4 &	41.5 &	-51.7	 & CORALIE \\
~58411.591838	& 20749.7 &	32.7 &	-169.6	 & CORALIE \\
~58411.648374	& 20664.0 &	28.6 &	-73.4	 & CORALIE \\
~58417.594656	& 20602.4 &	43.3 &	-274.9	 & CORALIE \\
~58419.531857	& 20635.3 &	31.1 &	-71.7	 & CORALIE \\
~58426.579955	& 20998.7 &	34.5 &	-88.5	 & CORALIE \\
~58465.662441	& 20607.2 &	20.7 &	-63.9	 & CORALIE \\
~58478.592577	& 21109.9 &	14.5 &	-42.7	 & CORALIE \\
~58479.593054	& 21206.5 &	15.5 &	-47.2	 & CORALIE \\
~58487.570334	& 21766.1 &	18.0 &	-71.7	 & CORALIE \\
~58500.538320	& 22047.5 &	28.3 &	-32.1	 & CORALIE \\
~58408.66413 & 	20733.1  &	10.8  &	-35 	& FEROS \\
~58411.74045 & 	20675.5  &	9.3  &	-29 	& FEROS \\
~58412.64109 & 	20671.7  &	8.3  &	-52 	& FEROS \\
~58413.54882 & 	20791.6  &	7.8  &	-29 	& FEROS \\
~58414.63269 & 	20630.9  &	8.3  &	-35 	& FEROS \\
~58415.63086 & 	20635.0  &	9.7  &	-4 		& FEROS \\
~58416.59652 & 	20608.9  &	8.3  &	 15 	& FEROS \\
~58418.59589 & 	20662.7  &	8.3  &	-40 	& FEROS \\
~58419.56945 & 	20659.0  &	8.6  &	-14 	& FEROS \\
~58423.65772 & 	21140.9  &	8.7  &	-57 	& FEROS \\
~58424.57114 & 	20896.8  &	8.4  &	-26 	& FEROS \\
~58428.72879 & 	21898.1  &	9.5  &	-15 	& FEROS \\
~58430.68560 & 	21856.6  &	8.8  &	 15 	& FEROS \\
~58450.63188 & 	21655.5  &	8.6  &	-17 	& FEROS \\
~58451.56707 & 	21856.7  &	9.3  &	 55 	& FEROS \\
~58451.58556 & 	21762.3  &	8.1  &	-35 	& FEROS \\
~58452.57292 & 	21747.4  &	8.8  &	-9 		& FEROS \\
\hline                                 
\end{tabular}
\end{table}

\begin{table}
\caption{\label{tab:157rvs}Radial velocity measurements from CORALIE and FEROS for TOI-157. The two FEROS RVs were not included in the global modelling of the system.}
\centering                          
\begin{tabular}{l c c c c}        
\hline\hline                 
BJD & RV  & $\sigma_{\mathrm{RV}}$ & BIS  & Instrument \\
(- 2,400,000) & (\ms) & (\ms) &  (\ms) &  \\
\hline
~58394.715066 &	-8782.2 &	118.7 &	-130.7	&	CORALIE	\\
~58397.866695 &	-8941.8 &	72.6 &	-59.6	&	CORALIE	\\
~58414.670583 &	-8868.6 &	52.6 &	15.3	&	CORALIE	\\
~58417.691914 &	-8498.7 &	62.6 &	-35.8	&	CORALIE	\\
~58418.770424 &	-8855.5 &	55.2 &	-14.9	&	CORALIE	\\
~58419.814201 &	-8543.7 &	53.7 &	-36.5	&	CORALIE	\\
~58427.857757 &	-8658.9 &	53.5 &	-7.8	&	CORALIE	\\
~58433.659427 &	-8876.8 &	75.0 &	-105.2	&	CORALIE	\\
~58455.780420 &	-8742.2 &	44.4 &	-120.3	&	CORALIE	\\
~58456.769193 &	-8860.9 &	90.3 &	-117.4	&	CORALIE	\\
~58457.708950 &	-8681.1 &	73.9 &	38.3	&	CORALIE	\\
~58458.705351 &	-8951.0 &	79.1 &	-124.8	&	CORALIE	\\
~58460.665537 $\dagger$ &	-9183.8 &	211.2 &	414.9	&	CORALIE	\\
~58461.712853 &	-8534.8 &	57.9 &	-40.7	&	CORALIE	\\
~58462.651604 &	-8907.4 &	58.5 &	-3.4	&	CORALIE	\\
~58463.574905 &	-8503.6 &	59.6 &	-12.2	&	CORALIE	\\
~58463.809868 &	-8526.8 &	50.3 &	-130.3	&	CORALIE	\\
~58464.562176 &	-8926.0 &	51.5 &	106.0	&	CORALIE	\\
~58464.749963 &	-8958.1 &	65.2 &	64.1	&	CORALIE	\\
~58467.599544 &	-8571.1 &	61.5 &	-15.2	&	CORALIE	\\
~58471.590055 &	-8625.6 &	57.7 &	-90.1	&	CORALIE	\\
~58474.675323 &	-8691.5 &	50.8 &	44.9	&	CORALIE	\\
~58475.657076 &	-8538.2 &	43.6 &	139.4	&	CORALIE	\\
~58486.705154 &	-8543.7 &	54.0 &	-47.5	&	CORALIE	\\
~58381.883623 &	-8663.9	& 16.3	& 25	& FEROS \\
~58383.881139 &	-8672.1	& 14.9	& 61	& FEROS \\
\hline                                 
\end{tabular}
\tablefoot{$\dagger$ Low S/N RV-measurement from BJD 58460.665537, not included in global analysis.}
\end{table}

\begin{table}
\caption{\label{tab:169rvs}Radial velocity measurements from CORALIE and FEROS for TOI-169.}
\centering                          
\begin{tabular}{l c c c c}        
\hline\hline                 
BJD & RV  & $\sigma_{\mathrm{RV}}$ & BIS  & Instrument \\
(- 2,400,000) & (\ms) & (\ms) &  (\ms) &  \\
\hline
~58411.74975   & 43526.9  &  14.0 	&  68 & FEROS \\
~58414.79857   & 43754.4  &  16.3 	& -66 & FEROS \\
~58418.63748   & 43562.4  &  12.1 	&  17 & FEROS \\
~58419.65549   & 43671.5  &  10.6 	& -28 & FEROS \\
~58423.68635   & 43716.1  &  11.5 	&  23 & FEROS \\
~58428.71918   & 43688.5  &  12.7 	&  15 & FEROS \\
~58429.67408   & 43529.5  &  11.2 	&  5 & FEROS \\
~58430.75367   & 43716.2  &  11.5 	& -7 & FEROS \\
~58450.72036   & 43696.4  &  10.4 	&  56 & FEROS \\
~58451.57625   & 43586.1  &  9.8 	&  29 & FEROS \\
~58648.913708 & 43623.24 & 63.7 &	119.9	& CORALIE \\
~58657.885639 & 43570.95 & 38.2 &	-27.9	& CORALIE \\
~58666.832642 & 43601.87 & 55.4 &	-66.4	& CORALIE \\
~58669.821282 & 43720.12 & 41.9 &	97.4	& CORALIE \\
~58677.860677 & 43523.81 & 53.0 &	-17.5	& CORALIE \\
~58679.934507 & 43490.82 & 42.3 &	18.9	& CORALIE \\
\hline                                 
\end{tabular}
\end{table}

\section{Light curves}
\begin{figure} 
   \centering   
  \includegraphics[width=\columnwidth,trim={0cm 0cm 0cm 0cm},clip]{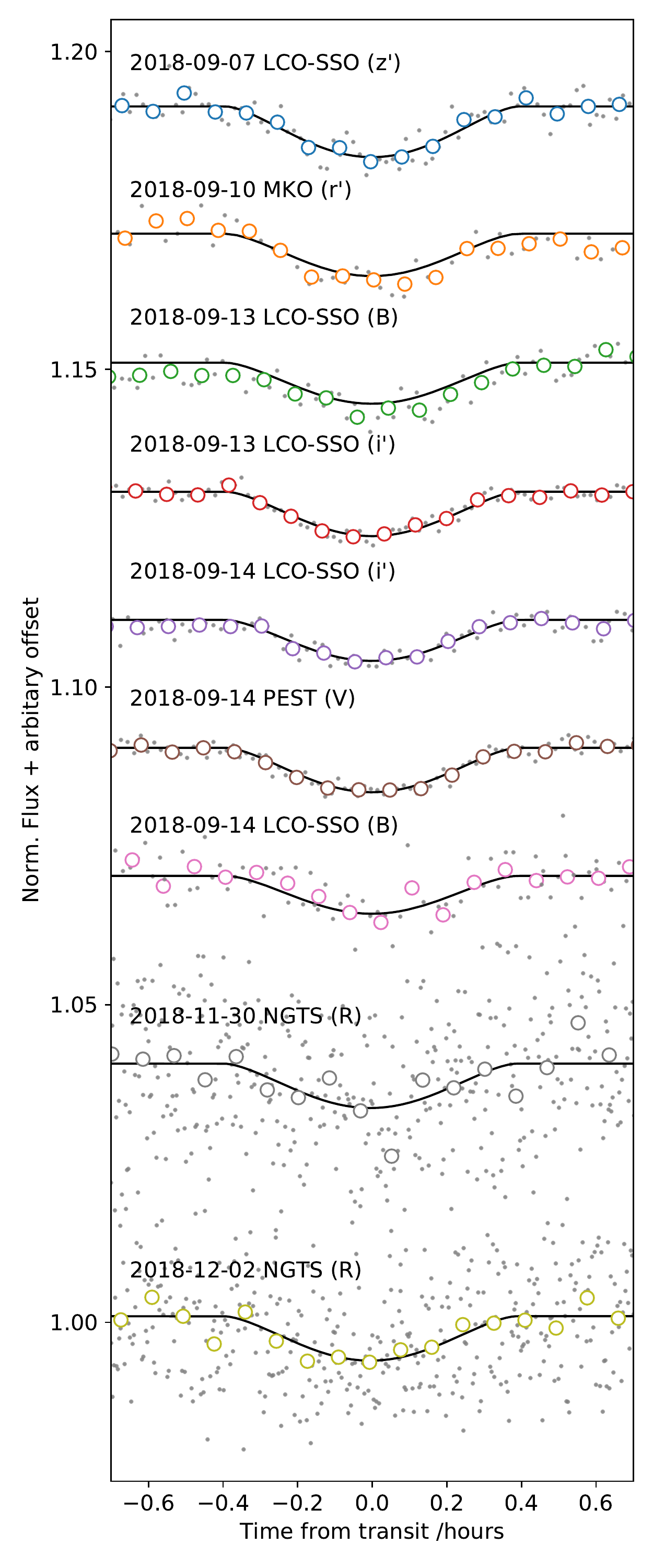}
      \caption{\label{fig:129_FUP} Ground based photometric follow-up data for HIP~65Ab from LCO-SSO, MKO, PEST and NGTS. The open circles are data binned to 5 min. }
\end{figure}  

\begin{figure} 
   \centering   
  \includegraphics[width=\columnwidth,trim={0cm 0cm 0cm 0cm},clip]{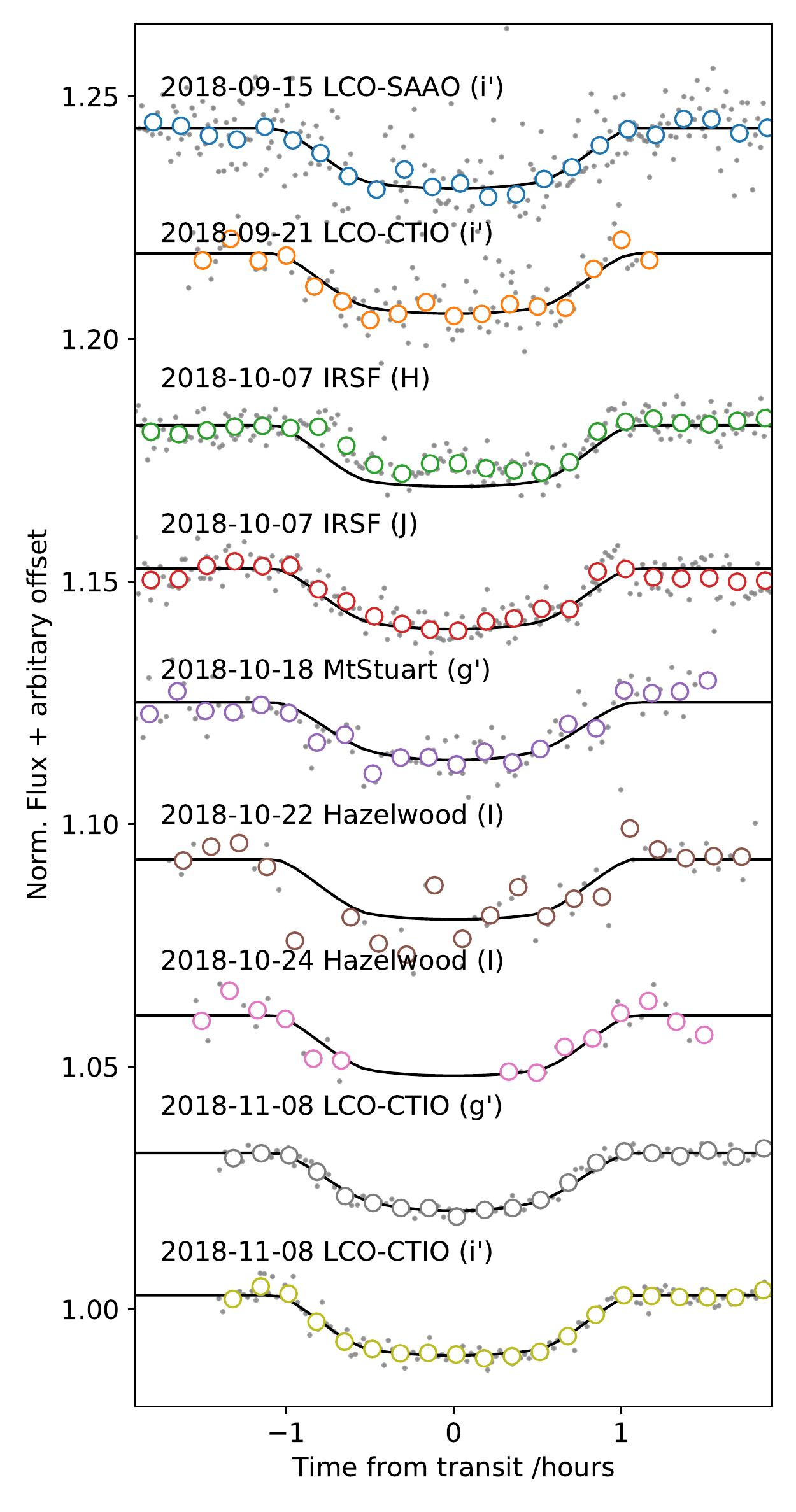}
  \includegraphics[width=\columnwidth,trim={0cm 0cm 0cm 0cm},clip]{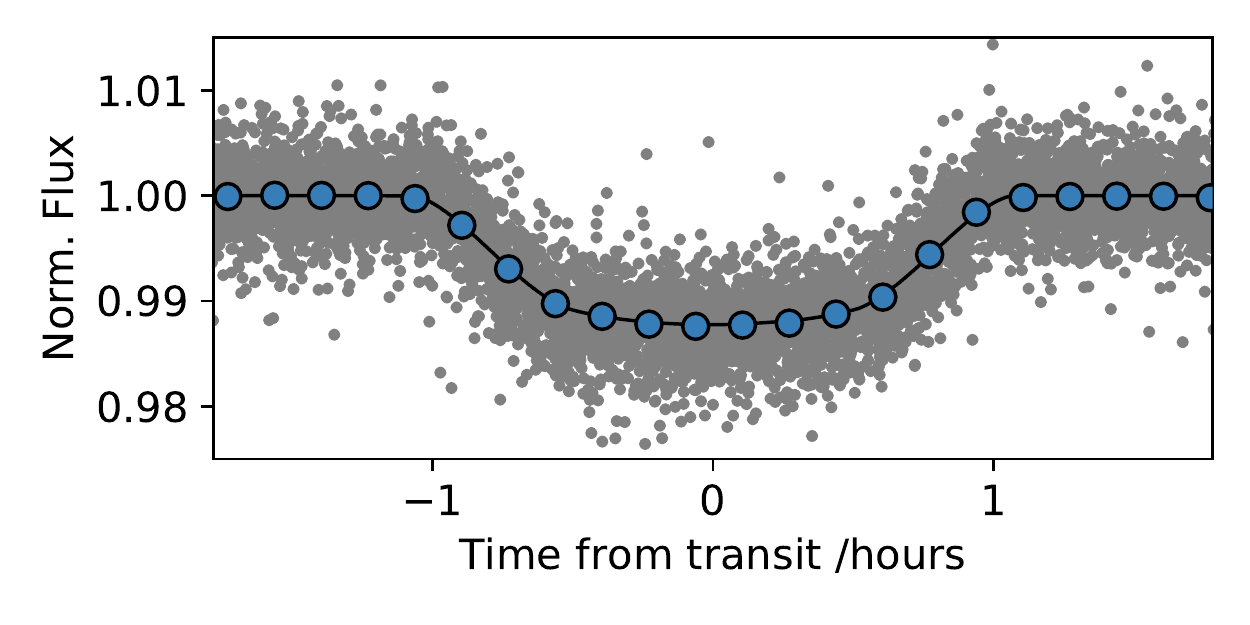}
      \caption{\label{fig:157_LC} \emph{Top:} Ground based photometric follow-up data for TOI-157b. The open circles are data binned to 10 min. \emph{Bottom:} Phase folded transit light curve for TOI-157b including \tess\ data and follow-up photometry in grey. The blue circles are the same data binned to 10 min. }
\end{figure}

\begin{figure} 
   \centering   
  \includegraphics[width=\columnwidth,trim={0cm 0cm 0cm 0cm},clip]{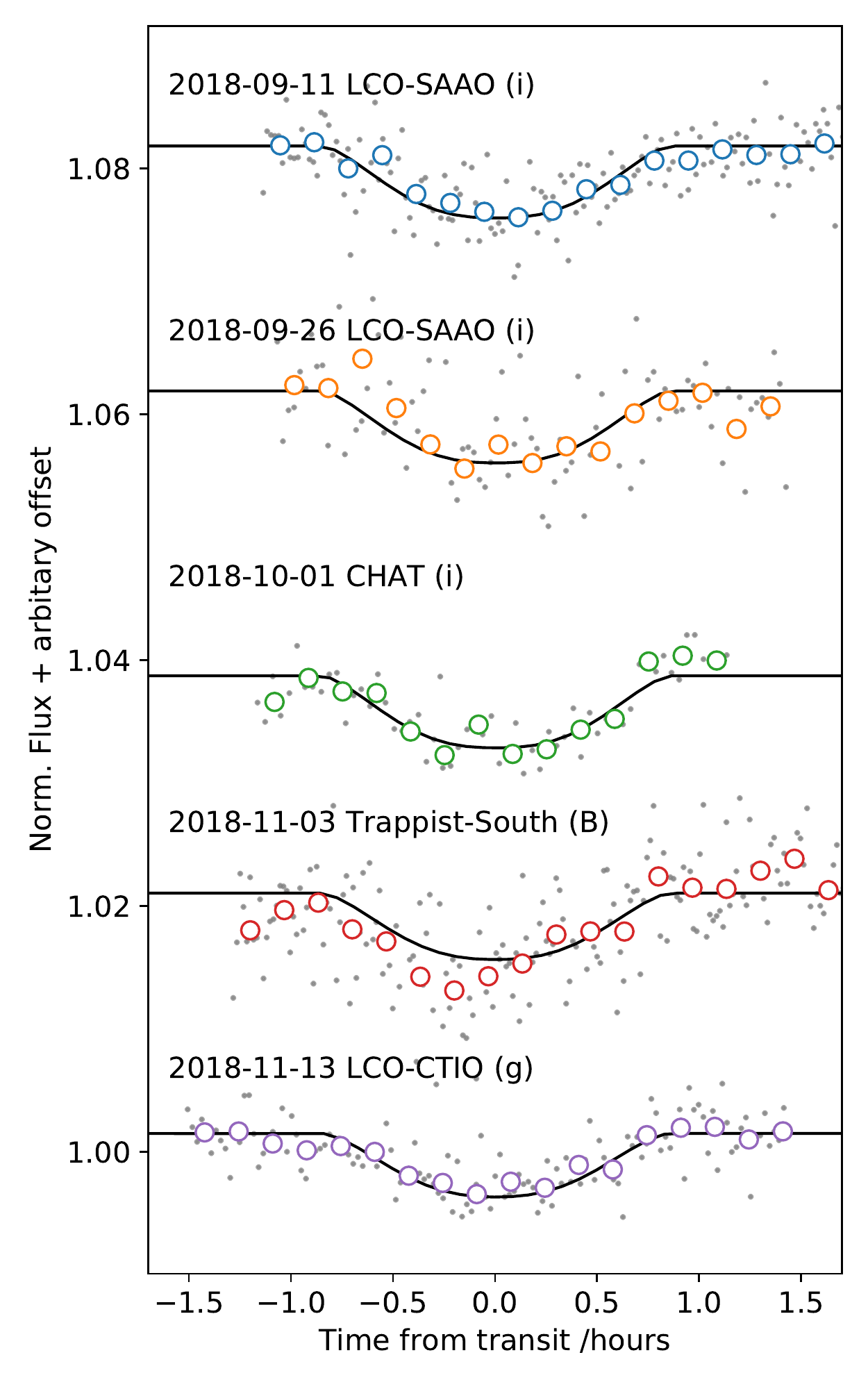}
  \includegraphics[width=\columnwidth,trim={0cm 0cm 0cm 0cm},clip]{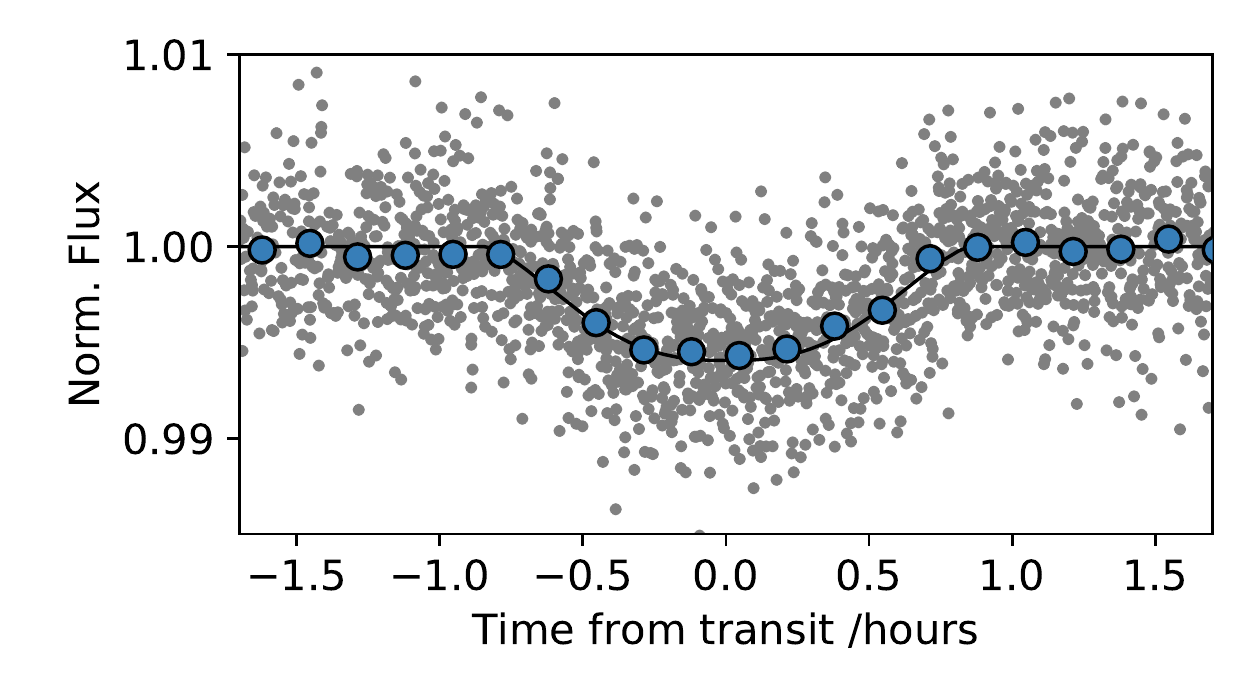}
        \caption{\label{fig:169_LC} \emph{Top:} Ground based photometric follow-up data for TOI-169 The open circles are data binned to 10 min.  \emph{Bottom:} Phase folded transit light curve for TOI-169 including \tess\ data and follow-up photometry, also with 10 min bins over-plotted as blue circles.}
\end{figure}

\end{appendix}
\end{document}